\documentclass[a4paper,11pt]{extarticle}  

\usepackage[T1]{fontenc}

\usepackage{stylesheet}

\usepackage[left=1in,top=1in,bottom=.65in,right=1in]{geometry}

\usepackage[english]{babel}

\usepackage{longtable}

\usepackage{subcaption}

\newcommand{\ts}[1]{\textsuperscript{#1}}

\newcommand{\credi}[0]{{\sc credi}}
\newcommand{\sdi}[0]{{\sc solar credi}}
\newcommand{\wdi}[0]{{\sc wind credi}}

\title{The Climatological Renewable Energy Deviation Index (\credi)}

\author{%
\textbf{%
Laurens P. Stoop \orcidlink{0000-0003-2756-5653},\textcolor{Accent}{\textsuperscript{1,2,3,*}} %
Karin van der Wiel \orcidlink{0000-0001-9365-5759},\textcolor{Accent}{\textsuperscript{4}} %
William Zappa \orcidlink{0000-0001-6810-7224},\textcolor{Accent}{\textsuperscript{3}} %
Arno Haverkamp \orcidlink{0000-0001-6947-7892},\textcolor{Accent}{\textsuperscript{3}} %
Ad J. Feelders \orcidlink{0000-0003-4525-1949},\textcolor{Accent}{\textsuperscript{1}} %
Machteld van den Broek\orcidlink{0000-0003-1028-1742}\textcolor{Accent}{\textsuperscript{5}} }\\[0.5em]
\begin{small}%
\textcolor{Accent}{\textsuperscript{1}}Information and Computing Science, Utrecht University, the Netherlands \\ 
\textcolor{Accent}{\textsuperscript{2}}Copernicus Institute of Sustainable Development, Utrecht University, the Netherlands\\ 
\textcolor{Accent}{\textsuperscript{3}}TenneT TSO B.V., Arnhem, the Netherlands\\ 
\textcolor{Accent}{\textsuperscript{4}}Royal Netherlands Meteorological Institute (KNMI), the Netherlands\\ 
\textcolor{Accent}{\textsuperscript{5}}Delft University of Technology, Faculty Technology, Policy, and Management, the Netherlands\\[0.5em] 
\textcolor{Accent}{\textsuperscript{*}}Corresponding Author: \textcolor{Accent}{laurens.stoop@tennet.eu} \\ \end{small}
}

\date{\today}

\begin{document}

\thispagestyle{empty}

\begin{center}

\parbox{420pt}{This is the author accepted manuscript (AAM) of the following published article:}  \vspace{20pt}

\begin{tabular}{|r|l|}\hline
\textbf{DOI} & \url{https://doi.org/10.1088/1748-9326/ad27b9} \\ \hline
\textbf{arXiv DOI} & \url{https://doi.org/10.48550/arXiv.2307.08909} \\ \hline
 & \\ \hline
\textbf{author(s)} & \parbox{290pt}{Laurens P. Stoop, Karin van der Wiel, William Zappa, Arno Haverkamp, Ad J. Feelders, Machteld van den Broek} \\ \hline 
\textbf{title} & \parbox{290pt}{The Climatological Renewable Energy Deviation Index (\credi) } \\ \hline
\textbf{publication date} & February 2024 \\ \hline
\textbf{journal} & Enviromental Research Lettes \\ \hline
\textbf{volume} & 19 \\ \hline
\textbf{page numbers} & 034021 \\ \hline
\end{tabular}

\vspace{30pt}
\parbox{420pt}{This AAM version corresponds to the author's final version of the article, as accepted by the journal. However, it has not been copy-edited or formatted by the journal. \\[20pt]
This AAM is deposited under a Creative Commons Attribution--ShareAlike (CC-BY-SA) license.} 
\end{center}

\newpage

\pagenumbering{arabic}

\maketitle

\begin{abstract}
We propose an index to quantify and analyse the impact of climatological variability on the energy system at different timescales.
We define the Climatological Renewable Energy Deviation Index (\credi) as the cumulative anomaly of a renewable resource with respect to its climate over a specific time period of interest.
For this we introduce the smooth, yet physical, hourly rolling window climatology that captures the expected hourly to yearly behaviour of renewable resources.
We analyse the presented index at decadal, annual and (sub-)seasonal timescales for a sample region and discuss scientific and practical implications.\\
\credi{} is meant as an analytical tool for researchers and stakeholders to help them quantify, understand, and explain, the impact of energy-meteorological variability on future energy system. 
Improved understanding translates to better assessments of how renewable resources, and the associated risks for energy security, may fare in current and future climatological settings.
The practical use of the index is in resource planning. 
For example transmission system operators may be able to adjust short-term planning to reduce adequacy issues before they occur or combine the index with storyline event selection for improved assessments of climate change related risks.
\end{abstract}

\vspace{1pc}
\noindent{\it \color{Highlight} Keywords}: Resource droughts, Resource Adequacy, Renewable Energy Drought, Dunkelflaute, Wind Drought
\vspace{1pc}


\section{Introduction}
The energy system is changing. 
This is due to the increased deployment of renewable energy generators, like wind turbines and solar panels; changes in electricity demand, from increased use of heat pumps and electric vehicles; and climatic changes influencing the weather dependent parts of the system. 
It is crucial to understand the full dynamics of the (future) energy system, both for policy making and energy security reasons~\autocite{craig2022disconnect}.

Knowing the impact of and link between the energy system and weather-related variability on daily to inter-annual and decadal timescales is vital for robust design and planning of future energy systems~\autocite{Bloomfield2021nextgen,craig2022disconnect,McKenna2022}. 
Meteorological variability leads to temporal variability. 
Not only in renewable energy production, but also in energy demand, changing the way energy systems have to be operated and controlled~\autocite{craig2022disconnect}.

Energy system models are vital to capture the impact of this variability~\parencite{gernaat2021climate}. 
However, their complexity results in high computational burdens that grows exponentially with the simulation period~\parencite{wuijts2022modelchar,Price2022,craig2022disconnect,wuijts2023linking,Grochowicz2023}. 
Incorporating large climate datasets that capture energy-meteorological variability in \emph{operational hourly} energy system models is thus, as of yet, unfeasible~\parencite{Harang2020,craig2022disconnect,wuijts2023linking}. 
Even so, understanding the scale of this variability, can aid  system operators in their task to ensure both short- and long-term energy security~\parencite{craig2022disconnect,wuijts2023linking,Hu2023}. 
Therefore, alternative approaches are needed to assess energy-meteorological variability~\parencite{craig2022disconnect,Dubus2022PECD}. 
While a number of methods exists to model and/or select challenging high impact events using basic statistical principles~\parencite[e.g.][]{vanderwiel2019extreme,Ohlendorf2020,OteroFelipe2021,stoop2021detection,vanderMost2022,Boston2022,Hu2023,Allen2023}, we aim to define a physics based and intuitive to understand metric to quantify energy-meteorological variability across timescales.

In developing this metric, we were inspired by the hydrological sciences. 
For drought monitoring, a number of indices have proven useful for both scientific assessment and operational use. 
These drought indices, such as the Climatological Water Balance~\parencite[CWB;][]{gleick1985regional,Gleick1986}, the Standardised Precipitation Index \parencite[SPI;][]{mckee1993}, and the Standardised Precipitation-Evapotranspiration Index \parencite[SPEI;][]{VicenteSerrano2010}), are based on precipitation deficits (anomaly of precipitation, or anomaly of the difference between potential evapotranspiration and precipitation) and are used to assess the temporal development of dry or wet periods. 
Furthermore, they have been used to assess the influence of inter-annual to multi-decadal variability, and of climate change on the temporal variability of hydrological drought~\parencite[e.g.][]{Quiring2009,Stagge2015,Cammalleri2021,vanderWiel2022}. 
As \textcite{Allen2023} showed, some aspects of these indices and their use in assessing hydrological variability can be transferred to the energy-meteorological domain.
However, where \textcite{Allen2023} developed direct analogues of the SPI and SPEI metrics using probabilistic descriptions, we took inspiration from aspects of these metrics, their use in operational applications, and combined this with the need for physically grounded storylines in energy system operation.

\begin{figure}[ht]
        \centering
        \includegraphics[width=\textwidth]{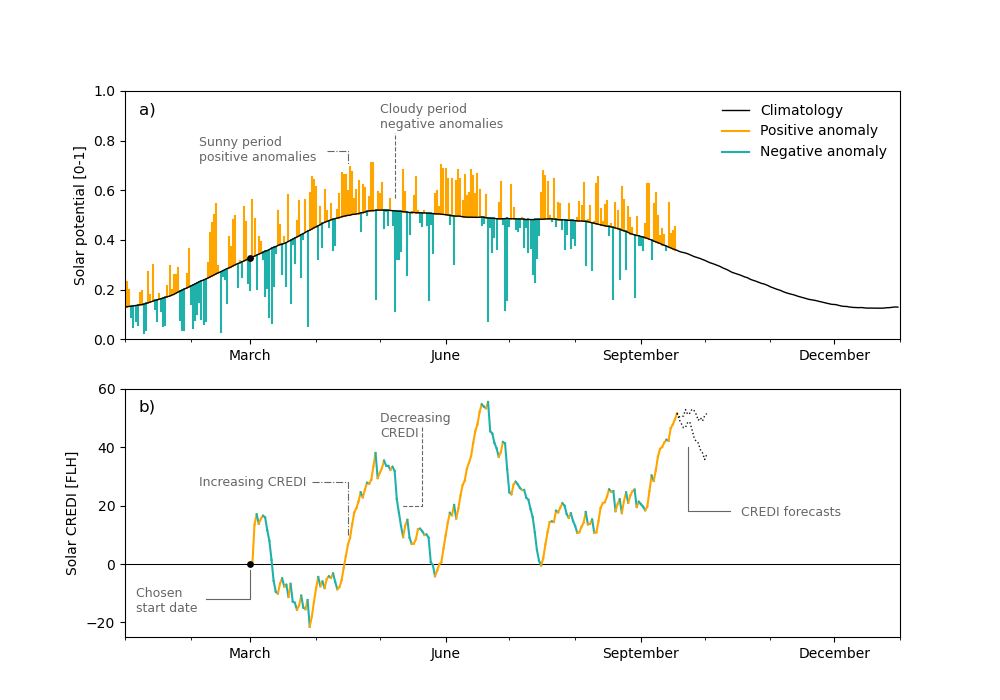}
        \caption{
                Illustration of climatological renewable energy deviation index. 
                Given the climate of a renewable resource (black line in panel~a), the instantaneous anomaly can be calculated (orange/teal bars in panel~a). 
                Positive anomalies (orange) increase, and negative anomalies (teal) decrease the index, which starts at zero at the start of period of analysis (panel~b). 
                The illustration shows solar potential anomalies for 2021 with respect to a 1991-2020 climate, and the \sdi{} with a starting point at 1~March. 
                Two meteorological forecast ensemble members converted to \credi{} are shown to indicate a use-case for grid-operators. }
        \label{fig:overview}
\end{figure}

We define the Climatological Renewable Energy Deviation Index (\credi) as the cumulative anomaly of a renewable resource with respect to its climate over a specific time period of interest (Figure~\ref{fig:overview}). 
Given this definition, this study addresses the following considerations: (a) how do you define the climatic behaviour of a highly variable renewable resource, like wind or solar? and (b) how do you analyse the \credi{} score at different timescales; like (sub-)seasonal, annual, or multi-decadal?

The paper is structured as follows. 
In Section~\ref{secCP2:clima}, we define the hourly rolling window climate and the index. 
In Section~\ref{secCP2:data}, we indicate the data used.
In Section~\ref{secCP2:analysis}, we analyse the index at different timescales and discuss the best starting point. 
In the Section~\ref{secCP2:discussion} we discuss our definition of the index. 
Finally, in Section~\ref{secCP2:conclusion}, a synthesis of our findings is presented and potential use cases in research and/or operational application are outlined. 
Supporting Information (SI) with additional figures and observational analysis is available online.


\section{Definition of the climatic characterisation and index}\label{secCP2:clima}

Within the atmospheric sciences the climate of a region is defined as the statistical-mean weather conditions prevailing in that region~\parencite{Arguez2011}. 
The \textcite[WMO;][]{wmo2017normals} has a standardised method for calculation of the \emph{climatological normals}, which comes down to calculating monthly or daily mean values over a 30-year period. 
The climate, or mean expected behaviour, of renewable resources could be defined similarly. 
However, monthly or daily climatological values are not suitable due to the highly variable nature of renewable resources like wind and solar energy, and the need to balance the power grid at shorter timescales.

We can distinguish four relevant timescales that cover the main modes of energy-meteorological variability. 
Namely:
\begin{enumerate}
        \item annual to decadal timescales: variability caused by interactions in the coupled ocean-atmosphere-system, e.g. modes of variability like the El-Ni\~no-Southern Oscillation \parencite[ENSO;][]{ipcc6_wg1_variability} or the North Atlantic Oscillation \parencite[NAO;][]{Wanner2001},
        \item seasonal timescale: variability caused by the revolution of the Earth around the Sun and the directly related variation of the solar declination angle,
        \item sub-seasonal timescale: variability caused by the cumulative interplay at various timescales, associated with the passing of weather systems and the changes in their persistence and occurrence,
        \item daily timescale: variability caused by the revolution of the Earth around its axis, and the directly related times of sunrise, sunset, and the solar elevation angle.
\end{enumerate}
When studying the generation potential of wind or solar, all these timescales of variability should be considered.


\subsection{A climatology of renewable resources and the use of hourly rolling windows}

\begin{figure}[ht]
        \centering
        \includegraphics[width=.95\textwidth]{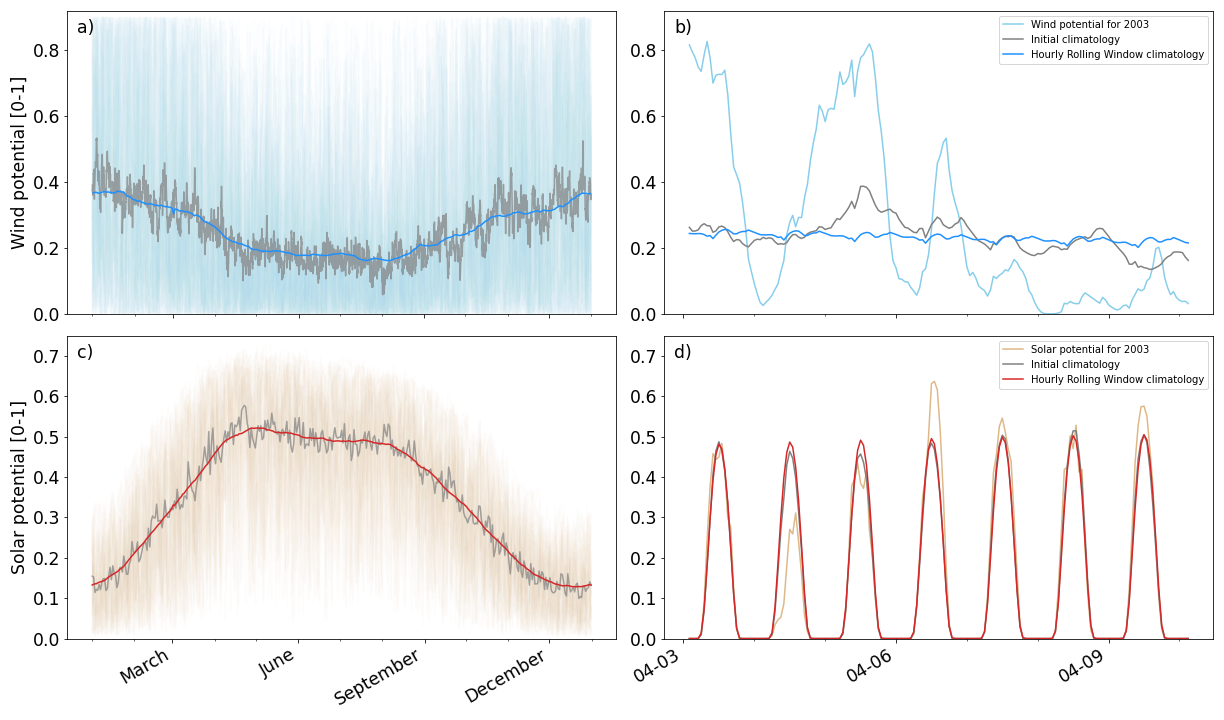}
        \caption{
                Comparison of different methods for computing the climate of the potential generation for wind (top), and solar (bottom), for the period 1991-2020. 
                Figures (a,c) show the hourly generation potentials for each year in this period (light blue for wind and orange for solar), the simple average-based climate (grey, see main text for details) and the hourly rolling window climate (blue and red, for wind, solar, respectively). 
                Figures (b,d) show the same, but specifically for the period 3-10 April 2003. 
                For clarity only 13:00 for each day of the year is shown in Figure (c).}
        \label{fig:climate}
\end{figure}

The highly variable nature of the wind and solar resources makes that a straightforward 30-year daily mean does not result in a useful definition of their climate (see SI Section~\ref{SIA:climatology}). 
The same holds for an initial estimate by averaging each ordinal hour over 30 years (Figure~\ref{fig:climate}a,c). 
Though this simple average-based climatology does capture the mean expected behaviour on annual timescales, the random fluctuations from day-to-day and hour-to-hour cannot be explained by physical processes in this climatological definition. 
To remove these random fluctuations more data would be needed to obtain the desired, physical, smooth , but physical, climatology. 
However, considering a period longer than 30-years is ineffective, as climate change would start to influence the result\autocite{wmo2017normals}. 
Applying a simple running mean to this simple average-based climate timeseries is undesirable, as that would remove the diurnal cycle, which has a physical origin and is of large importance for our application in the energy sector.

We therefore define an \emph{hourly rolling window} climate, meaning that we first group the same time of day, and then, for each 'hour-of-the-day'-group, we apply a 30-year running mean (see SI Section~\ref{app:clima_hourly}). 
The hourly rolling window climate ($C$) of a renewable resource potential $P$ for hour-of-the-year $h$ is computed by:
\begin{equation}
        C_P(h) = \frac{1}{n}\sum_{y=1}^n ~\sum_{h' \in \{h+24d\}_{d=-\Delta}^{d=+\Delta}} ~ \frac{P(y, h')}{2 \Delta +1} ,\hspace{1cm} h = 1,2,\ldots,8760 \label{eq:HRWclim}
\end{equation}
where $n$ is the number of years, $h$ is the hour of the year from 1 to 8760, $\Delta$ is half the window size (days) and $P(y,h')$ is the generation potential for hour $h'$ of year $y$. 
In line with \parencite{Arguez2011} an unweighted average and $n=$~30 years are used.
See Figure~\ref{fig:climate} for a comparison between the different methods.

It should be noted that two details where omitted in formula \ref{eq:HRWclim}. 
First, the hour-of-the-year is cyclic in nature, meaning that the first hour of year $y$ follows the last hour of year $y-1$. 
While this is implemented, for reasons of clarity this is not included.
Second, to deal with leap years, we discard February 29\ts{th} when computing the climatology. 
The climatology of each hour of the day for February 29\ts{th} is then defined by the mean value of that hour of the day of February 28\ts{th} and March 1\ts{st}. 
This addresses the lack of data for 29~February and keeps a simple formalism.

The choice for the size of the rolling window is somewhat arbitrary. 
Sensitivity tests indicate that the window size should be bigger then 20~days to smooth any remaining nonphysical day-to-day variability, but smaller then 60 days to avoid over-smoothing the annual cycle (SI Section~\ref{app:Harmonics}). 
Within this range the exact size of the window does not affect the use of the index. 
Here, we choose a window size of 40~days.

By using the hourly rolling window climate, both the importance of the various timescales and the need for more data points to get a smooth climatological function are addressed. 
It is essential that the climatological definition used in the calculation of the deviation index for wind or solar energy is physical (i.e. does not contain random fluctuations), such that anomalies represent variability due to the weather, decoupled from the climate.


\subsection{The Climatological Renewable Energy Deviation Index (\credi)}

We define the \credi{} to be the cumulative anomaly of a renewable resource with respect to its climate over a specific time period of interest from a chosen starting point in two steps (Figure~\ref{fig:overview}). 
First, we determine the anomaly of a renewable resource, as the difference between the hourly generation potential of that resource and its climate (i.e. its expected value), taken from the computed hourly rolling window climate. 
Second, from an initial chosen starting point we sum these anomalies over a time period of interest.

More formally: let $P(y,h)$ denote the generation potential for ordinal hour $h$ of year $y$, and let $C_P(h)$ denote the climate for ordinal hour $h$ for that potential $P$. 
The anomaly $A_C(y,h)$ of a renewable resource for ordinal hour $h$ of year $y$ is then defined as:
\begin{equation}
        A_C(y,h) = P(y,h) - C_P(h).
\end{equation}

The \credi{} over a given period of time is defined as the cumulative sum (or running total) of $A_C$ over that period. 
For example, if we align the starting point with the start of the year, the \credi{} on the $i$-th hour of that year ($y$) is:
\begin{equation}
        \mbox{\credi}(y,i) = \sum_{h=1}^{i} A_C(y,h), \hspace{2cm} i = 1,2,\ldots,8760
\end{equation}

When interpreting the index, the following should be considered. 
A change in \credi{} over time is an indication of either an excess or deficit of the renewable resource potential with respect to its climatic normal (Figure~\ref{fig:overview}b). 
A stable \credi{} over a period indicates nominal renewable resource potential with respect to its climate. 

Specifically, the \credi{} score has the unit Full Load Hours (FLH) and at a given time informs the user of the cumulative surplus or deficit generation potential over the period considered with respect to its nominal behaviour.
So given a fixed time window, the distribution of the \credi{} score calculated then provides insight into the properties of a connected storage unit, like the  (dis-)charge potential.
FLHs depend on the installed capacity, therefore if the installed capacity of a resource is known or assumed, the index allows for direct assessment of the storage volume and power needed to always generate nominally within the fixed time window used.

For clarity, when the index is applied to a specific resource, we first refer to the resource before the index acronym is given. 
For example, the \wdi{} refers to an assessment of the \credi{} of wind energy potential, and similarly for solar.


\subsection{The use of storylines in analysing \credi }\label{secCP2:storylines}
The index can be used to assess the temporal development of anomalous renewable energy generation. 
In line with the application of hydrological drought indices, a physical storylines approach~\parencite{Shepherd2019,vanderWiel2021} could be used. 
This approach can use regional climate change information while avoiding the strict limitations of a normal confidence-based approach applied in climate science. 
Storylines can be used to gain more insight into the driving processes, identify event analogues, and investigate similar events in alternative energy systems or under future climate conditions. 
Utilising these insights in, for example, resource adequacy assessments or system design studies, will likely lead to a more robust energy system.

Selection of relevant events can be based on historical adequacy assessments (like the \parencite{tennet2023} Adequacy Outlook). 
As shown by \textcite{vanderwiel2019extreme,vanderWiel2021}, event analogues can then be found in large energy-climate datasets that incorporate climate change~\parencite{craig2022disconnect,Dubus2022PECD}. 
By studying these analogues the physical processes and likelihood of these events can be assessed.

To demonstrate the index at different timescales and to highlight relevant considerations in the application of the \credi, we selected the years 1996, 1998, 2003 and 2016 as storylines. 
The year 1996 was chosen specifically, as one of the most challenging years for resource adequacy in the Netherlands and Germany in a future net-zero emission energy system~\parencite[][p.56]{tennet2023}. 
In the analysis of the potential for hydrogen generation from wind, 2003 and 2010 where found to be anomalously low~\parencite[][p.58-61]{tennet2023}. 
Both 1998 and 2016 where chosen as they represent the most anomalous years of the index for solar and wind, respectively.


\section{Data}\label{secCP2:data}
We used the preliminary 4th version of the Pan-European Climate Database to demonstrate the \credi{} in this paper~\parencite[PECDv4.0;][]{Dubus2022PECD}. 
This database, developed by Copernicus Climate Change Services (C3S) in cooperation with the European Network of Transmission System Operators for Electricity (ENTSO-E) will be the new standard database used for all common Transmission System Operator (TSO) studies. 
The full database will be openly available as part of the new C3S-Energy dataset, expected in late 2023 (\url{https://climate.copernicus.eu/energy/}). 

To showcase the developed index all figures show data from the preliminary PECDv4.0 of the northern region of the Netherlands.
This region is the NUTS statistical region `NL1' and covers the provinces of Groningen, Friesland and Drenthe, see: \url{https://en.wikipedia.org/wiki/NUTS_statistical_regions_of_the_Netherlands}.
While the region is named `NL01' in the PECD dataset, the NUTS-code is used here.
While we focus in the main paper on the NL1 region, in the supplement we show additional regions reflecting some of the diversity within Europe.
Further details in SI Section~\ref{SIF:dataext}.



\section{Application of the \credi{} at different timescales}\label{secCP2:analysis}
In this section we show the application of the index at decadal, seasonal and sub-seasonal timescales in the context of modelling future energy systems.
The considerations associated with choosing a starting point for the \credi{} calculation is especially relevant at (sub-)seasonal timescales, and will be discussed.

On daily timescales the weather is extremely variable, but it depends on local conditions and short-term battery storage comes into play~\autocite{parzen2023value}. 
For most regions the maximum cost-effective storage based on the surplus charging capacity from wind and/or solar is in the order of 8 hours to 4 days~\parencite{Livingston2020,Sepulveda2021,parzen2023value}. 
For these reasons, we make no assessment on daily timescales here. 
However, due to the relevance of short-term events for the energy system, an example of a 8-day study window in \credi{} is provided.


\subsection{Annual to decadal variability in \credi}\label{sc:decadal}
At annual to decadal timescales the index can be used to assess the impact of large scale oscillations in the ocean and atmosphere on the availability of a renewable energy resource. 
These long-term deviations from the climate are relevant, e.g., because they offer sources of meteorological predictability \parencite{Hawkins2009,Scaife2014}, or because stakeholders look at 10~year time periods to estimate return of investments~\parencite{tennet2023}.

\begin{figure}[ht]
        \centering
        \includegraphics[width=\textwidth]{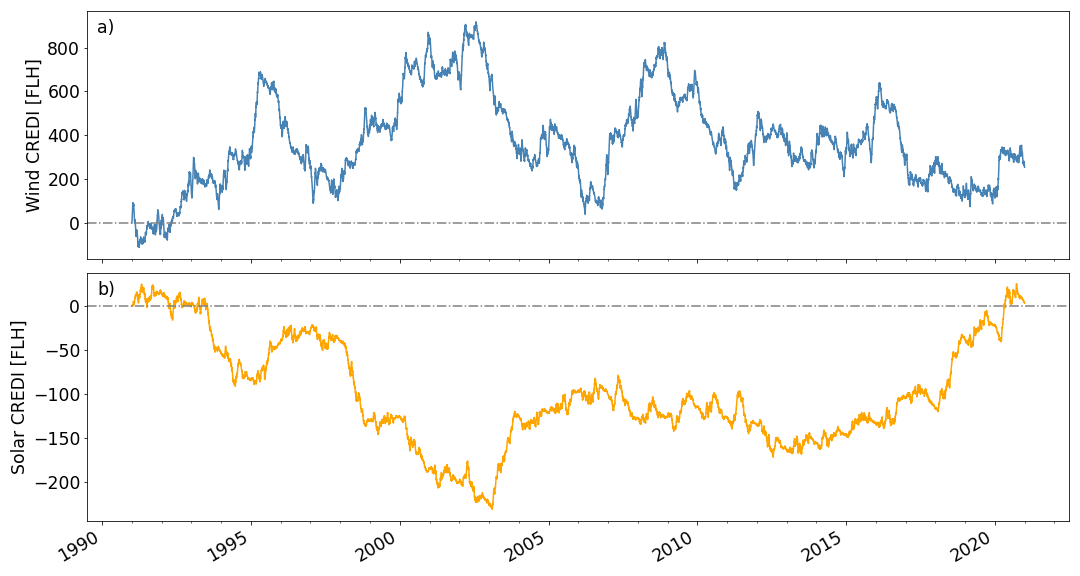}
        \caption{
                Hourly Wind (a) and Solar (b) \credi{} over the period 1991-2020 for `NL1'. 
                As the climate was calculated over the same period, by definition the \credi{} sums to zero over the full period.}
        \label{fig:analysis_decadal}
\end{figure}

Over the past 30~years, large inter-annual variation is observed in the \wdi{} (Figure~\ref{fig:analysis_decadal}a). 
The cumulative effect of variations at seasonal scales resulted in higher than expected wind generation potential from 1991 to 2002, while from 2010 onwards \wdi{} declined indicating lower than expected wind generation potential. 
These general variations are in line with those found by~\textcite{wohland2019significant,stoop2021detection}. 

Similarly, the \sdi{} shows inter-annual variability. 
From 1991 to 2003 \sdi{} shows a general decrease, indicating less than average potential generation from solar. 
Within this period, a strong reduction in the periods 1993-1995 and 1998-2002 is observed (Figure~\ref{fig:analysis_decadal}b). 
In the period 2005-2018, \sdi{} is flat, showing that the solar potential was as expected from climate. 
After this period a steady increase in the \sdi{} is observed, indicating higher than expected potential generation.

The values of \sdi{} are generally lower than those of the \wdi{}. 
This is directly related to the diurnal cycle, which by definition gives zero solar potential at night and low values in the morning and evening. 
Consequently, the sum of the anomalies over a given period is smaller than for wind potential, which has values for all 24~hours in a day.

Finally, while the impact of the relative observed variability depends on the ratio of installed capacities, we observe that the inter-annual energy-meteorological variability is mainly driven by the wind resource in the analysed region (i.e., the northern of the Netherlands). 
And though the {\sc wind} and {\sc solar} \credi s show strong anti-correlated behaviour during some years (e.g. from 1991 to 2002), in others this is not the case (e.g. from 2004 to 2005). 
At decadal timescales, wind and solar balance the system somewhat, but they are not suited to fully negate the variability of their counterpart.


\subsection{Seasonal variability in \credi}\label{sc:seasonal}
When assessing the seasonal energy-meteorological variability using the \credi, the starting point determines the way the temporal development of the index is perceived. 
In line with definitions of hydrological drought, the starting point determines the separation between energy surplus (wet) and deficit (dry) years. 
As the index is intended to capture the energy-meteorological variability, the start date is picked such that the biggest range if \credi{} at the end of, and throughout, the year is observed. 

Comparing \credi{} starting points for each month of the year, we found that these should \emph{not} be the same for wind and solar (SI Section~\ref{SIB:startdate}). 
We use May 1\ts{st} as the starting point for wind, as it gives the widest distribution of the index at the end of the analysis window in this particular region.  
For solar no clear distinction is found between a December or January starting point, we chose to use January 1\ts{st} here.

\begin{figure}[hb]
        \centering
        \includegraphics[width=\textwidth]{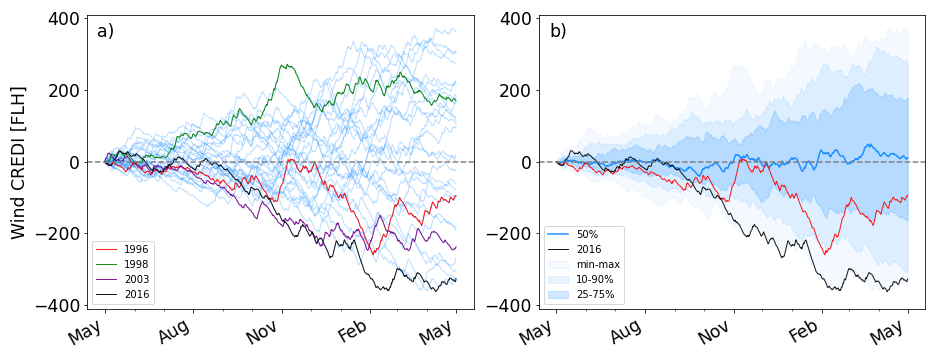}    
        \caption{
                Hourly \wdi{} per analysis year over the period May 1991 to April 2021 for `NL1'. 
                Figure a) shows the specific progression of \wdi{} for each year (blue lines). 
                Figure b) shows the distribution of the \wdi{} for each hour of the year, namely the 50\ts{th} percentile (blue line), the 25-75, 10-90 percentile and min-max range (shaded blue, see legend). 
                Four exemplary storylines are shown, namely 1996 (red), 1998 (green), 2003 (purple) and 2016 (black).
        }
        \label{fig:analysis_seasonal_wind}
\end{figure}

For the yearly \wdi, it is obvious that an individual year can either be anomalously positive or negative, and that variations throughout a year are large (Figure~\ref{fig:analysis_seasonal_wind}a). 
This results in a wide range of yearly storylines. 
The 25-75\% spread of the index grows to $\pm180$ FLH over a year (Figure~\ref{fig:analysis_seasonal_wind}b). 
The most extreme negative year in the period considered for \wdi{} was 2016. 
In that year, from about September onwards, the wind potential was almost consistently below expected with $350$ less FLHs at the end of the analysed period.

As an example of the use of \wdi{} for storyline analysis we look at 1996. From May to October the index is relatively flat, indicating that the wind potential was as expected from its climate (red line in Figure~\ref{fig:analysis_seasonal_wind}b). 
Then, a strong reduction is observed in the \wdi{} from December to the end of January, indicating much lower then average potential generation from wind. Part of this deviation is compensated by higher than normal generation potential in February of 1997.

\begin{figure}[ht]
        \centering
        \includegraphics[width=\textwidth]{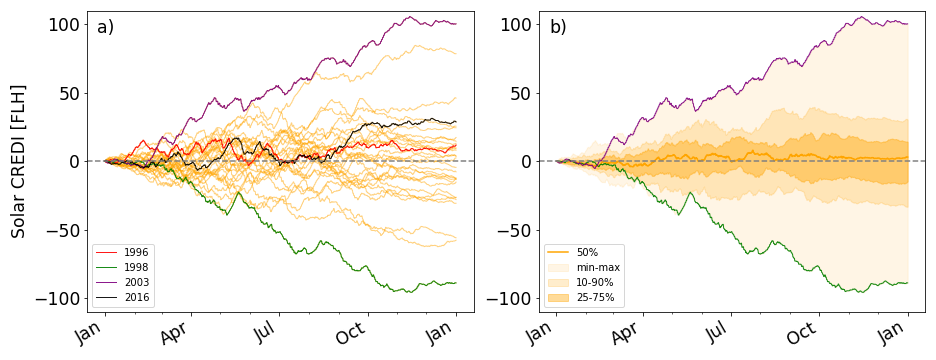} 
        \caption{
                Hourly \sdi{} per year over the period 1991-2020 for `NL1'. 
                As shown in Figure~\ref{fig:analysis_seasonal_wind}, but the \sdi{} is shown in orange hues.
        }
        \label{fig:analysis_seasonal_solar}
\end{figure}

As noted earlier, values of yearly \sdi{} are smaller than of \wdi{} (Figure~\ref{fig:analysis_seasonal_solar}a), with an average spread (25-75\%) of $\pm18$~FLHs, and uncommon spread (10-90\%) of $\pm35$~FLHs spread over a year (Figure~\ref{fig:analysis_seasonal_solar}b). 
This indicates that \textasciitilde18~FLHs of total energy is needed to cover the deficit of the installed solar capacity in 50~\% of years and \textasciitilde35~FLHs to cover 80\% of years (Figure~\ref{fig:analysis_seasonal_solar}). 

The most extreme year of high solar potential was 2003; the most extreme year of low solar potential was 1998. 
Especially 2003 is remembered for its extremely warm and sunny summer \parencite{GarcaHerrera2010}.

\subsection{Sub-seasonal variability in \credi}\label{sc:subseasonal}

At sub-seasonal timescales, similar to seasonal, the start point determines the way the temporal development of the index is perceived. 
We use 'energy'-seasons to capture the large scale changes on sub-seasonal timescales. 
For wind we define two seasons of interest: September to March, and April to August. 
For brevity, only the results found for wind in the winter `energy'-season are shown here, see SI Section~\ref{SIC:seasonal} for the other and solar. 
Alternative definitions of 'energy'-seasons can be relevant, especially for regions that have different sub-seasonal behaviour then the `NL1'-region shown here.

\begin{figure}[t]
        \centering
        \includegraphics[width=\textwidth]{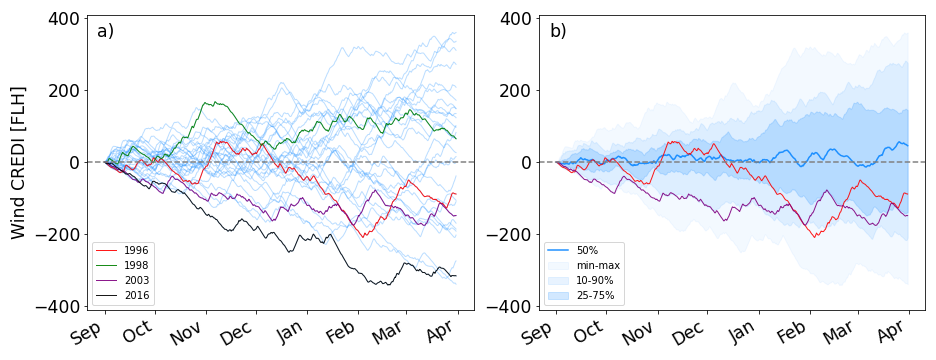} 
        \caption{
                Hourly winter \wdi{} per season (Sep.-Apr.) for 1991-2021 for `NL1'. 
                Figure a) shows the specific progression of \wdi{} for each summer season (blue lines). 
                In addition, four example storylines are represented, namely those starting in 1996 (red), 1998 (green), 2003 (purple) and 2016 (black). 
                Figure b) shows two storylines (1996, 2003) and the hourly distribution of the \wdi{}, namely the 50\ts{th} percentile (blue line), the 25-75, 10-90 percentile, and min-max range (shaded blue, see legend).
        }
        \label{fig:analysis_sub-season-winter_wind}
\end{figure}

It is obvious that different years show quite different characteristics (Figure~\ref{fig:analysis_sub-season-winter_wind}a) and individual winter seasons can differ greatly. 
As expected, the sub-seasonal timescale is emphasised more. 
For instance, the anomalous index-development in 1996 described in Section~\ref{sc:seasonal} is more clearly visible. 
Especially the strong reduction in \wdi{} from December to the end of January stands out as a period of much lower than normal wind generation potential.

\subsection{A short-term study window for event-based \credi{}}\label{sc:shortterm}
Finally, short-term events, e.g. \emph{Dunkelflautes}, can pose significant risk to highly renewable energy systems \parencite{tedesco2023,Mockert2022arxiv,vanderwiel2019extreme,Li2022,Sundar2022}. 
A 8-day window for \credi{} aligns with previous work~\parencite{tennet2023}, and is investigated here, see SI Section~\ref{SID:shortterm} for additional figures and the top 50 8-day events. 

For short-term event analysis we do not pre-define the start point, all 8-day windows are considered. 
Overlapping events that share a five or more days, are removed from the analysis. 
While we only consider the lowest final \credi{} value for our event selection, other impact selection methods as described by \textcite{vanderWiel2021} can be used.

\begin{figure}[ht]
        \centering
        \includegraphics[width=\textwidth]{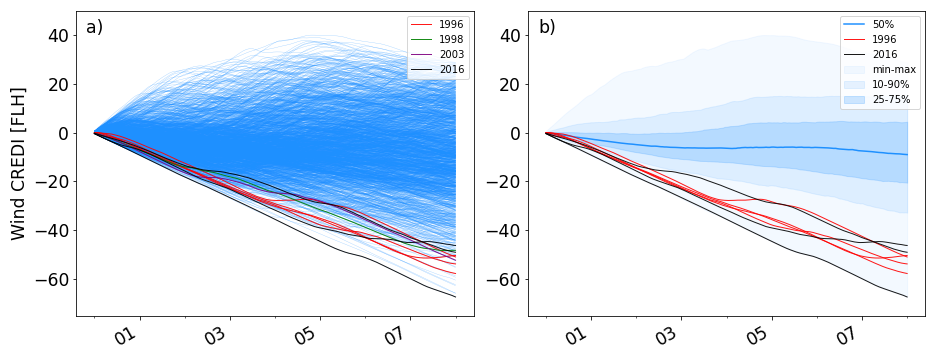}
        \caption{
                Hourly winter \wdi{} during 8-days for all events with less then 5 days overlapping in the period May 1991 to April 2021 for `NL1'. 
                The storylines show the analysis years 1996 (red, 4x), 1998 (green, 1x), 2003 (purple, 1x) and 2016 (black, 3x). 
                Further formatting as shown in Figure~\ref{fig:analysis_sub-season-winter_wind}.
        }
        \label{fig:analysis_short-term-winter_wind}
\end{figure}

Again we noted the large weather-caused variability between different 8-day periods (Figure~\ref{fig:analysis_short-term-winter_wind}a). 
The computed spread in Figure~\ref{fig:analysis_short-term-winter_wind}b considers events throughout the year. 
This can also be investigated on a seasonal basis for winter or summer-specific event information, or for shorter or longer events.  

The most extreme event is from 16-24 jan. 2017 and the analysis year 2016 is present 3 times in the top 50 events. 
While the specific 8-day event found in \textcite{tennet2023} is not the most extreme event, the analysis year 1996 does show up four times in the top 50 events. 
Indicating that the analysis year 1996 indeed stands out as quite exceptional.

\section{Discussion}\label{secCP2:discussion}
The presented index is defined as the cumulative anomaly of a renewable resource with respect to its climate. 
The method of determining the climate is thus vital and, as shown, should take into account the strong diurnal and annual cycle present in renewable energy resources. 
The calculation of the climate used here has a dependence on the size of the rolling window, which was primarily based on expert judgement. 
A longer timeseries, covering many decades, could be used for a cross-validation check to obtain the optimum \emph{rolling window} size, but the data source should be selected with great care, due to potential inconsistencies~\parencite{wohland2019significant,Wohland2022,Deser2023}. 
In previous work a climatic definition on harmonics has been effective~\parencite{Sabziparvar2014,Fischer2019,Rayson2021}, but we found it unsuitable here (see SI Section~\ref{app:Harmonics}). 

\credi{} should not be confused with the standardised energy indices recently introduced by \textcite{Allen2023}.
While we have been inspired by indices for monitoring hydrological droughts, their standardised energy indices are direct analogues. 
Meaning that those indices are a pure statistical assessment of the observed variance that rely heavily on the empirical distribution functions used \parencite[See Section~2][p.2-3]{Allen2023}.
However, in energy system operation and control, the specific sequence of observations and the deviation with respect to the expected patterns matter. 
The \credi{} presented incorporates these aspects.

When combined with weather forecasts, indices for hydrological drought can help policy makers make early decisions regarding societal risks~\parencite{Quiring2009,Stagge2015,Cammalleri2021,vanderWiel2022}. 
However, the operation of the electricity grid requires balance on very short timescales~\parencite{craig2022disconnect,tennet2023}. 
While we presented our index with an hourly resolution, further research is needed to investigate if the \credi{} can also be applied on these very short timescales. 
The examples provided, however, do already show \credi's usefulness in resilience planning, resource adequacy assessments, and as a metric for selecting events for robustness analysis.

In this introduction of the index, we applied it to the northern region of the Netherlands. 
However, as shown by \textcite{Pickering2020}, energy-meteorological variability is strongly region dependent. 
Therefore, the \credi{} should be calculated and analysed for each region separately. 
Due to the ease of application, and the intuitive analysis and interpretation of the index, this application to other regions is relatively straightforward (see SI Section~\ref{SIE:regions} for a few additional regions).


\section{Conclusion}\label{secCP2:conclusion}
Drawing inspiration from the work on drought monitoring indices, we have presented the hourly rolling window climatology and Climatological Renewable Energy Deviation Index (\credi{}). 
Given the relevance of both the diurnal and annual cycle in meteorology for energy applications, we recommend a simple but suitable definition of the background climate using an hourly rolling window approach. 
This new index is meant as an analytical method for researchers and stakeholders to help them understand and explain the impact of the variable nature of the weather on the energy system. 
The index computes the cumulative deviation or anomaly from the climatology for a chosen period.

The index can be used when understanding of energy-meteorological variability is key. 
For example, the \credi{} can be used as part of a resource adequacy analysis from TSOs to identify events which are likely to be a challenge in maintaining security of supply in a (future) power system driven by renewable energy sources. 
At the same time, the \credi{} could be used to assess the volume and power output of back-up resources needed for a given timescale, region, and energy system design.
Then, by using the event selection and analysis, as e.g. in \textcite{vanderWiel2021} for hydrological extremes, detailed event descriptions can be developed, systems can be stress tested, and further insight could be gained into energy-meteorological variability.


\section*{CRediT Author Statement}
Conceptualisation, Formal Analysis and Visualisation: LPS, Investigation, Methodology and Writing - Original Draft: LPS, KvdW, Writing - Review \& Editing: \emph{All listed authors}, Supervision and Funding acquisition: \emph{AJF, MvdB}. 

\section*{Acknowledgments}
Laurens P. Stoop received funding from the  Dutch Research Council (NWO) under grant number 647.003.005. 
The content of this paper and the views expressed in it are solely the author’s responsibility, and do not necessarily reflect the views of TenneT TSO B.V..

\section*{Open research} 
The implementation of the \credi{}, its use at different timescales, all code used to generate the figures, the data from the `NL1' region discussed and the full list of the most extreme short-term events found as presented in this study are available at Github via \url{https://github.com/laurensstoop/ccmetrics} with the MIT license. 

The full preliminary dataset of the PECDv4 containing the regional renewable resource potential for other technological definitions of wind and solar, or regions, then used in this study are not available due to ongoing validation. 
In due time the full PECDv4, including raw gridded and aggregated regional/national renewable resource potentials for a wide range of technological definitions, will be made available as part of the C3S Energy dataset and can be found through \url{https://climate.copernicus.eu/operational-service-energy-sector}. 
The framework describing the new C3S-Energy dataset, part of which is the PECDv4, can be found on: \url{https://climate.copernicus.eu/c3s2412-enhanced-operational-services-energy-sector}.

\printbibliography

\appendix
\beginsupplement

\newpage
\section{Comparison of Climatic Definitions of the Renewable Resources}\label{SIA:climatology}
Section 2 in the main text describes the four relevant timescales of energy-meteorological variability and shows the use of the hourly rolling window climate for renewable resources. 
Here we provide some additional detail on the observed variability of wind and solar potential (Section~\ref{app:variability}) and an analysis on the specific behaviour during each hour of the day (Section~\ref{app:clima_hourly}). 
We also highlight the use of different climate definitions (Section~\ref{app:Harmonics}) and discuss the sensitivity of the hourly rolling window climate on its window size (Section~\ref{app:sense}).


\subsection{Observed variability of wind and solar energy potential}\label{app:variability}
\begin{figure}[ht]
        \centering
        \includegraphics[width=\textwidth]{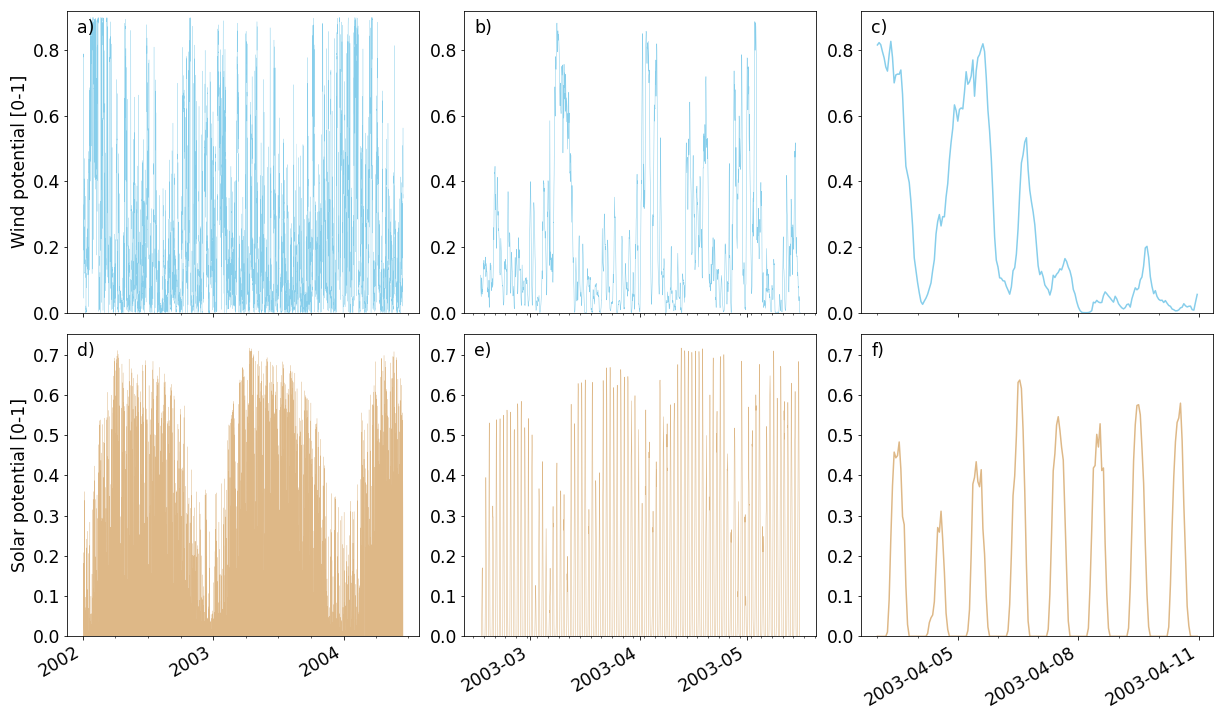}
        \caption{
                Timeseries of hourly generation potential of wind (top) and solar (bottom). 
                Showing variability on yearly (a,d), sub-seasonal (b,e) and daily (c,f) timescales. 
                This example shows data for the Dutch zone `NL1' from 2002-2004.}
        \label{fig:climatological_behaviour}
\end{figure}

Examples of typical behaviour of wind and solar energy are shown in Figure~\ref{fig:climatological_behaviour}. 
For wind, at seasonal timescales, we observe lower mean generation potential in the summer period (Figure~\ref{fig:climatological_behaviour}a, 2a.
when weather conditions are more stable. 
The higher and more variable generation potential in the autumn and winter period is associated with the quicker succession of storms in Europe (Figure~\ref{fig:climatological_behaviour}a,b).

For solar the difference between summer and winter is more pronounced, which is predominantly due to seasonal changes in the angle of declination of the sun (Figure~\ref{fig:climatological_behaviour}d and 2c
For both wind and solar the succession of large-scale high and low pressure systems can be observed (Figures~\ref{fig:climatological_behaviour}b,e). 
Additionally, as the efficiency of solar panels declines with increasing air temperature, a reduced solar generation potential is observed around noon after the summer solstice (Figure~\ref{SIfig:climate_solar_hourly}). 

On daily timescales the inherent diurnal cycle of the solar energy generation potential is very prominent and changes due to cloudiness are noticeable (Figure~\ref{fig:climatological_behaviour}f). 
For the wind generation potential no diurnal cycle is evident (Figure~\ref{SIfig:climate_wind_hourly}), but large intra- to multi-day changes associated with the passing of weather systems can clearly be observed (Figure~\ref{fig:climatological_behaviour}b,c).

The observed variability of wind and solar energy potential is in line with the large ensemble used by \textcite{vanderwiel2019extreme} and the decadal observations align with \textcite{Bett2013} and \textcite{wohland2019significant}.

It is clear that both wind and solar show strong variability at daily to yearly timescales (Figure~\ref{fig:climatological_behaviour}). 
To define a practically useful climate of the prevalent behaviour for the wind or solar energy resources, all these timescales of variability should be taken into account.


\subsection{Climatic characterisation for each hour of the day }\label{app:clima_hourly}
\begin{figure}[b]
    \centering
    \includegraphics[width=\textwidth]{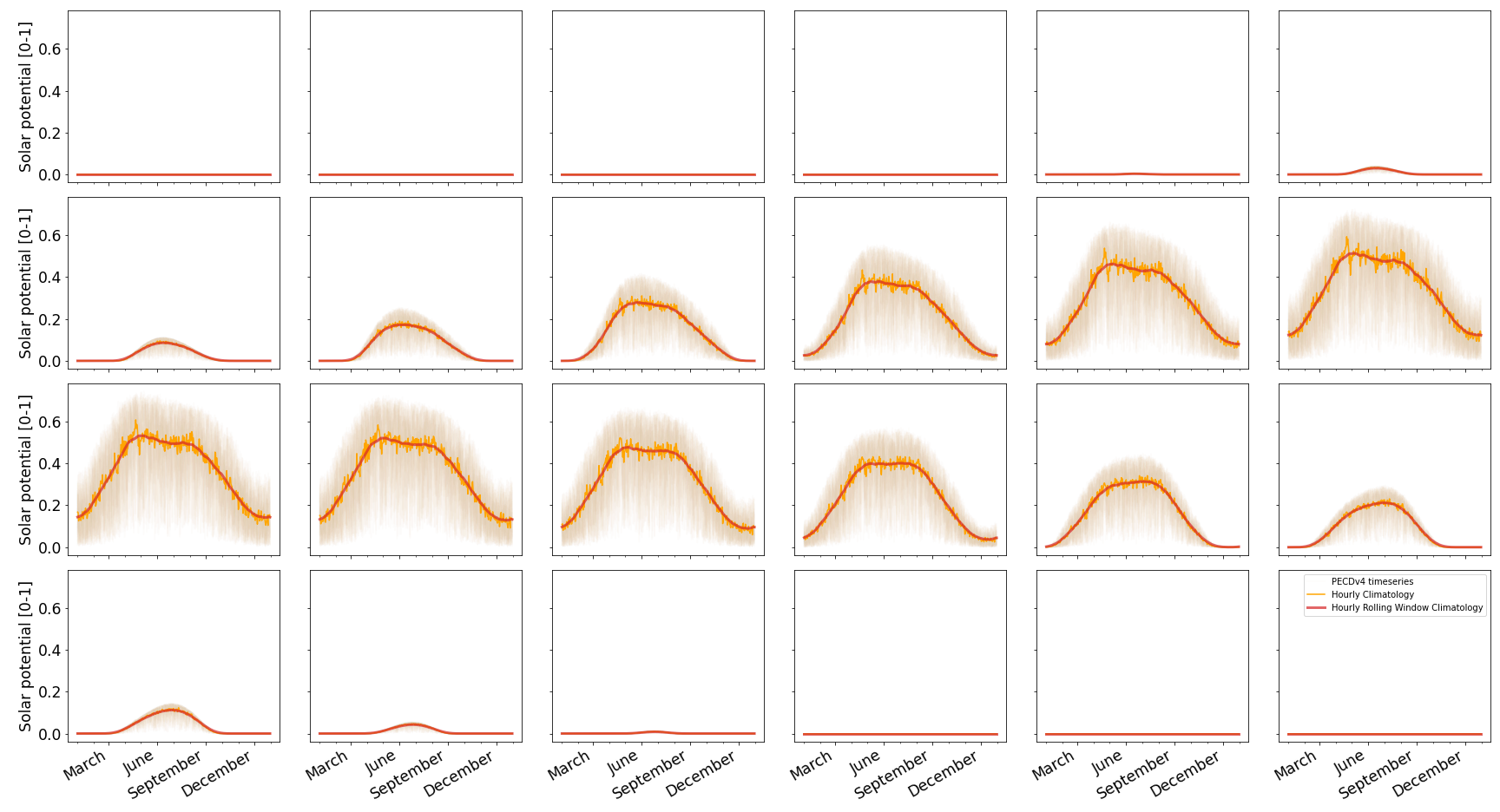}
    \caption{
    The climatological definitions for the solar potential generation is shown for each hour of the day over a year for the period 1991-2020 for `NL1'. 
    The figures show the simple average-based, initial, climate definition (grey), the hourly rolling window climate (red) and also include the full range of generation potentials in 1991-2020 (light orange). }
    \label{SIfig:climate_solar_hourly}
\end{figure}

\begin{figure}[ht!]
    \centering
    \includegraphics[width=\textwidth]{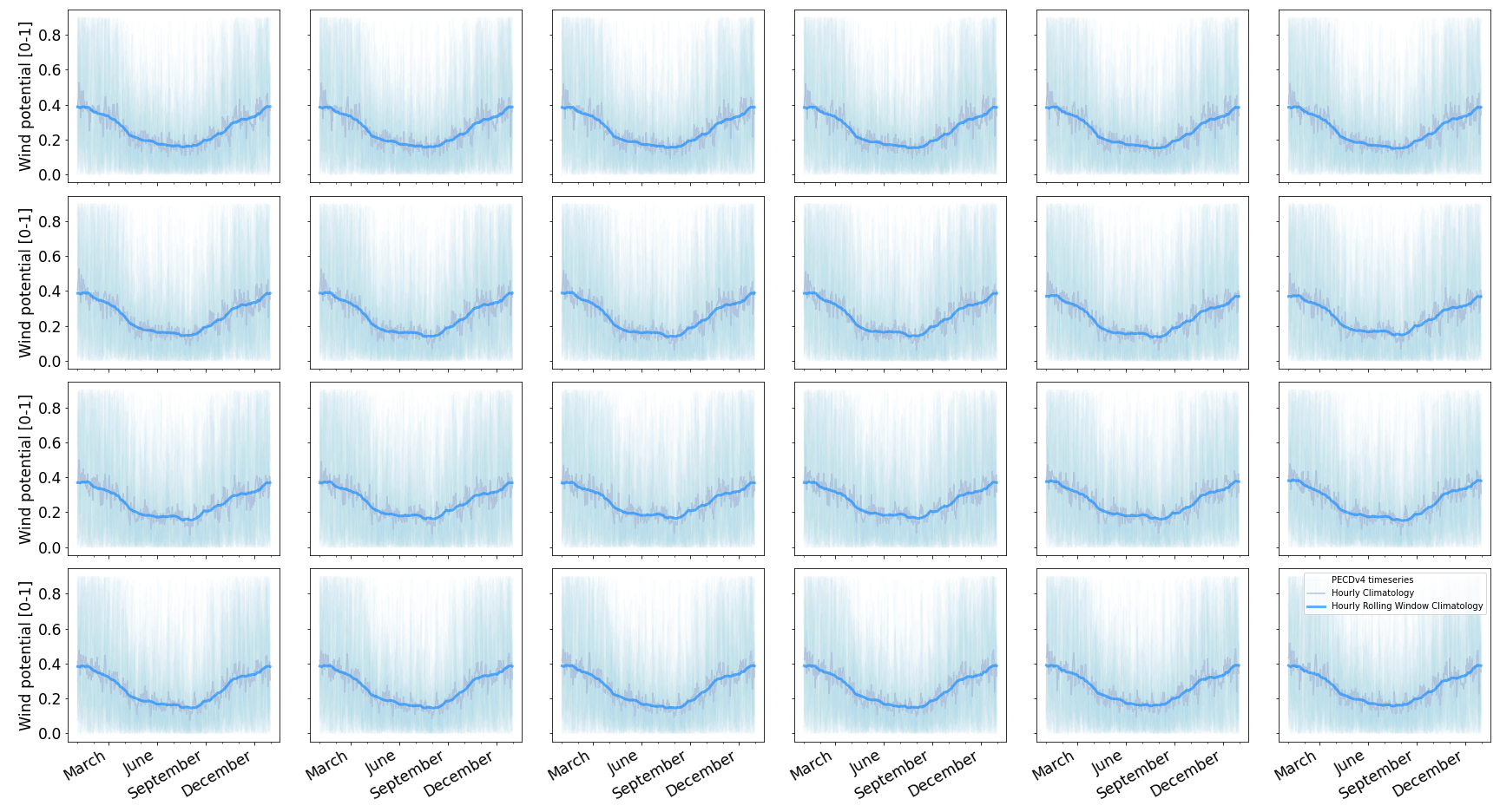}
    \caption{
    The climatological definitions for the wind potential generation is shown for each hour of the day over a year for the period 1991-2020 for `NL1'. 
    The figures show the simple average-based, initial, climate definition (grey), the hourly rolling window climate (dark blue) and also include the full range of generation potentials in 1991-2020 (light blue). }
    \label{SIfig:climate_wind_hourly}
\end{figure}

Section 2.1 in the main text describes the observed variability of wind and solar energy potential. 
Here we provide some additional figures (Figure~\ref{SIfig:climate_solar_hourly} and \ref{SIfig:climate_wind_hourly}) show the climatic behaviour throughout the year, for each hour of the day separately. 

For solar, the strong annual and diurnal cycle are very clearly visible. 
In addition, a few peculiarities can be observed related to how solar panels function. 
The efficiency of solar panels declines with increasing air temperature~\parencite{SaintDrenan2018}, leading to a reduced solar generation potential around noon after the summer solstice from the higher temperatures at this time of the year.  

For wind energy generation potential only the annual cycle of the seasonal variability of wind is clear and no clear distinction for the hour of the day can be made. 
The climatology for each hour of the day does not match perfectly and there are some minor differences observed.

\subsection{Comparison of climate definitions}\label{app:Harmonics}
\begin{figure}[b]
    \centering
    \includegraphics[width=0.9\textwidth]{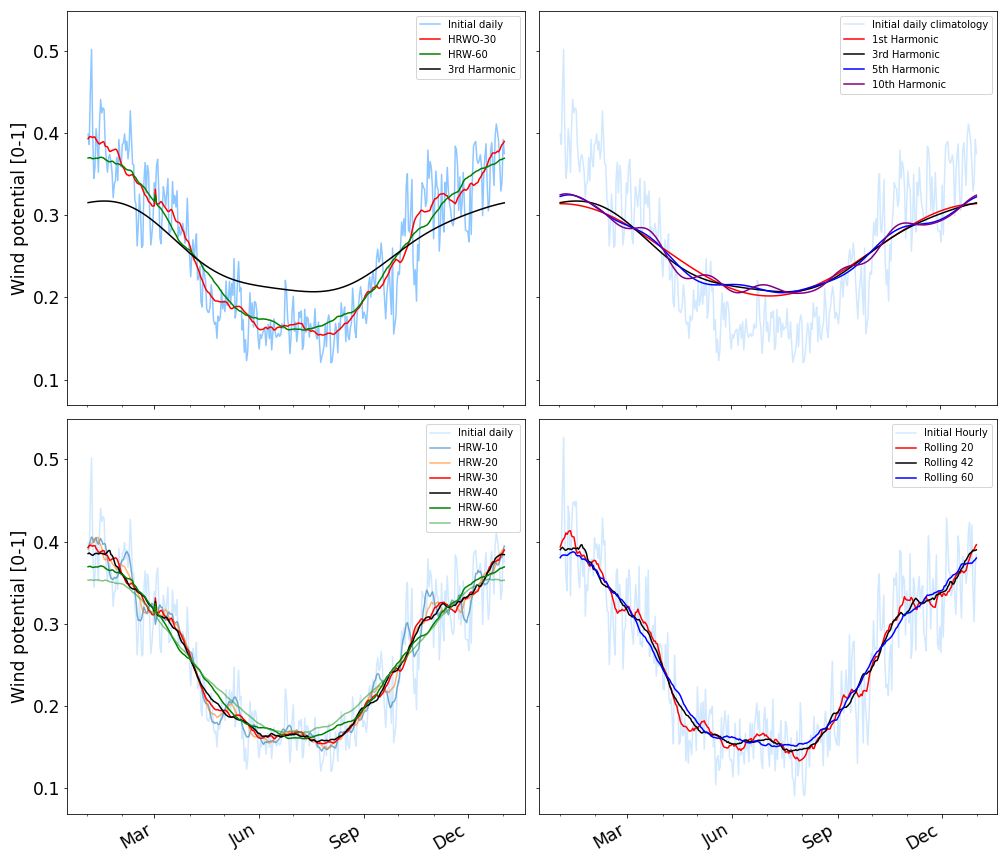}
    \caption{
    Comparison of windows for the hourly rolling window (HRW) climate for wind during the period 1991-2020 for `NL1'. 
    In light blue the yearly generation potentials from 1991 to 2020 are shown. 
    The simple average-based, initial, climate (grey, see main text for details) and various windows sizes (10,20,40,60,90,120 days) of the hourly rolling window climate (in purple, green, dark blue, yellow, black and orange, respectively) are shown.}
    \label{SIfig:clima_daily-harmonics_spv}
\end{figure}

\begin{figure}[t]
    \centering
    \includegraphics[width=0.9\textwidth]{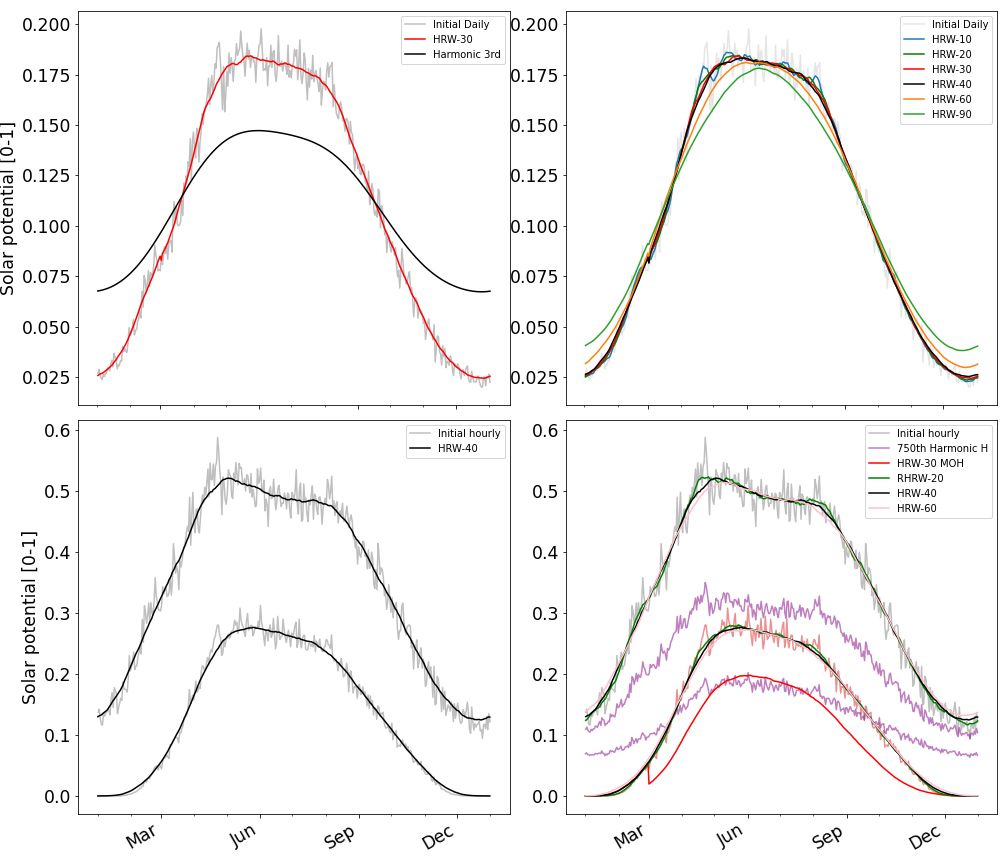}
    \caption{
    Comparison of windows for the hourly rolling window (HRW) climate for solar during the period 1991-2020 for `NL1'. 
    As shown in Figure~\ref{SIfig:clima_daily-harmonics_spv}, see legend for colours. }
    \label{SIfig:clima_daily-harmonics_won}
\end{figure}

Section 2.2 in the main text discusses the climate of a renewable resource. 
Here we provide some additional figures showing that both a daily~\parencite{wmo2017normals} and harmonic description of the climate are unsuitable for use in energy-meteorological applications (Figures~\ref{SIfig:clima_daily-harmonics_spv} \& \ref{SIfig:clima_daily-harmonics_won}). 
For the latter see the work of \textcite{Sabziparvar2014,Fischer2019,Rayson2021} for their use of the harmonic climate definition. 

\begin{figure}[ht!]
    \centering
    \includegraphics[width=0.9\textwidth]{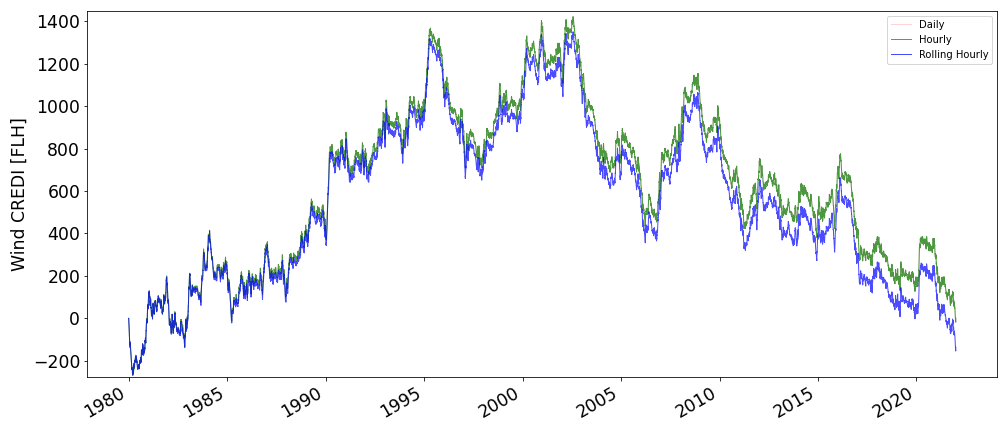}
    \caption{
    Comparison of the impact of a different climate definition on the resulting \credi{} for wind during the period 1991-2020 for `NL1'.}
    \label{SIfig:clima_impact}
\end{figure}

While the climate definitions are unsuitable, their impact on the \credi is limited (Figure~\ref{SIfig:clima_impact}). 

\clearpage

\subsection{Sensitivity of window size for Hourly Rolling Window climate definitions }\label{app:sense}
\begin{figure}[b]
    \centering
    \includegraphics[width=\textwidth]{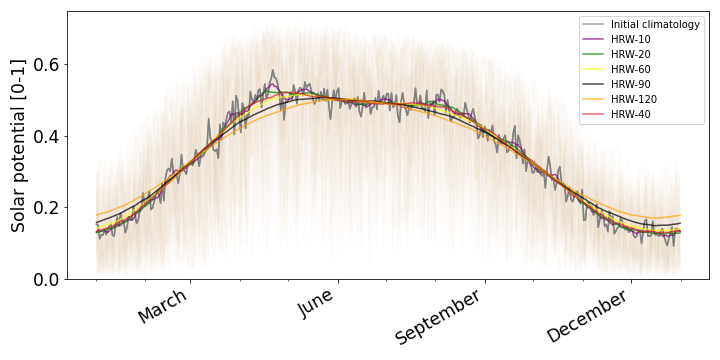}
    \caption{Comparison of windows for the hourly rolling window (HRW) climate for solar during the period 1991-2020 for `NL1'. In light orange the yearly generation potentials from 1991 to 2020 are shown. The simple average-based, initial, climate (grey, see main text for details) and various windows sizes (10,20,40,60,90,120 days) of the hourly rolling window climate (in purple, green, red, yellow, black and orange, respectively) are shown. For clarity only 13:00 for each day of the year is shown.}
    \label{SIfig:clima_sense_solar}
\end{figure}

\begin{figure}[t]
    \centering
    \includegraphics[width=\textwidth]{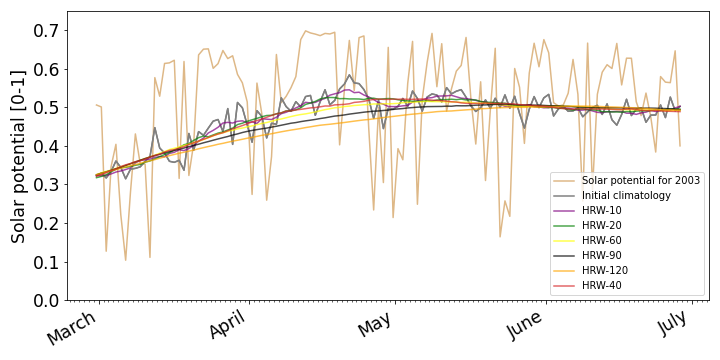}
    \caption{Comparison of windows for the hourly rolling window climate for solar, as shown in Figure~\ref{SIfig:clima_sense_solar}, but specifically for the period from March to June 2003.}
    \label{SIfig:clima_sense_zoom_solar}
\end{figure}

\begin{figure}[hb]
    \centering
    \includegraphics[width=\textwidth]{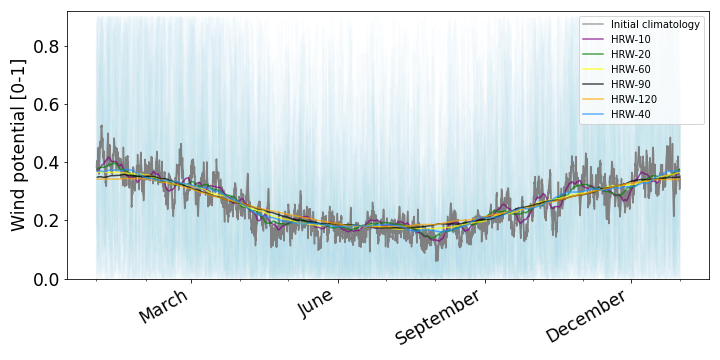}
    \caption{Comparison of windows for the hourly rolling window (HRW) climate for wind during the period 1991-2020 for `NL1'. In light blue the yearly generation potentials from 1991 to 2020 are shown. The simple average-based, initial, climate (grey, see main text for details) and various windows sizes (10,20,40,60,90,120 days) of the hourly rolling window climate (in purple, green, dark blue, yellow, black and orange, respectively) are shown.}
    \label{SIfig:clima_sense_wind}
\end{figure}

\begin{figure}[ht]
    \centering
    \includegraphics[width=\textwidth]{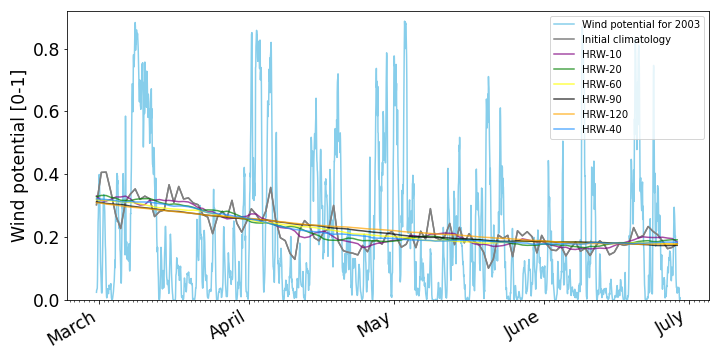}
    \caption{Comparison of windows for the hourly rolling window climate for wind for `NL1', as shown in Figure~\ref{SIfig:clima_sense_wind}, but specifically for the period from March to June 2003.}
    \label{SIfig:clima_sense_zoom_wind}
\end{figure}

A comparison of windows for the hourly rolling window climate is shown in Figure~\ref{SIfig:clima_sense_solar} \& \ref{SIfig:clima_sense_zoom_solar} for solar and Figure~\ref{SIfig:clima_sense_wind} \& \ref{SIfig:clima_sense_zoom_wind} for wind. 

For solar potential the hourly rolling window climate for a 10 day window is not suited as variations are observed on daily to weekly timescales that have no physical reason to be a recurrent over the years (see Figure~\ref{SIfig:clima_sense_solar}). Similar to the climate for wind, these fluctuations observed at the 10 day window would not constitute as a good definition of a climate. On the other hand, very large windows like those using the 60, 90 or 120 window, are very smooth throughout the year, underestimating for instance the peak of maximum solar potential near the end of April/start of May (see Figure~\ref{SIfig:clima_sense_zoom_solar}) and severely over estimating the winter dip in solar potential (Figure~\ref{SIfig:clima_sense_solar}). Again, inline with what was found with wind this indicates an over-smoothing of the yearly cycle and thus using these windows within the hourly rolling window climate would thus not be a good indicator of likely weather. A window size in the range of 20-60 days is adequate in capturing the persistent weather fluctuations and the annual peak solar potential, without underestimating the annual cycle.

For wind potential the hourly rolling window climate for a 10 day window is not suited as variations are observed on daily to monthly timescales that have no physical reason to be a recurrent over the years (see Figure~\ref{SIfig:clima_sense_zoom_wind}). As a climate is defined as the statistically-mean weather conditions \emph{prevailing} in a region, the short-term nature of the fluctuations observed at the 10 day window would not constitute as a good definition of a climate as the climate fluctuates on short timescales. The same holds for the 20 day window, albeit to a lesser extent. On the other hand, very large windows like those using the 90 or 120 window, are very smooth throughout the year. For most of the mid-winter period their climate is well below the initial, simple average-based, climate and during the summer above (see Figure~\ref{SIfig:clima_sense_wind}). This indicates an over-smoothing of the yearly cycle and thus using these windows within the hourly rolling window climate would thus not be a good indicator of likely weather. A window size in the range of 20-90 days is adequate in capturing the persistent weather fluctuations throughout the year, without underestimating the annual cycle.

\section{Annual start date analysis for \credi}\label{SIB:startdate}
\begin{figure}[t]
\centering
\begin{subfigure}[t]{0.32\linewidth}
    \includegraphics[width=\linewidth]{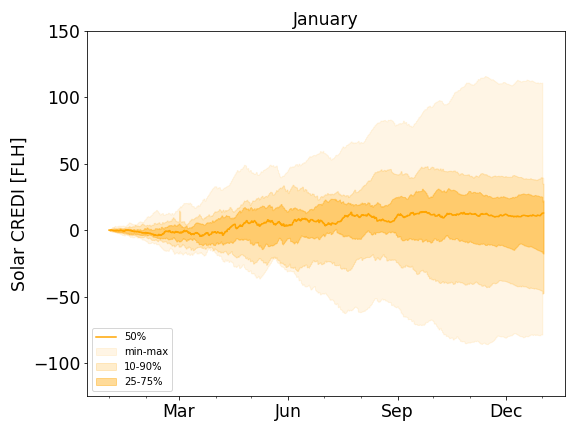}
    \caption{January }
\end{subfigure}
\begin{subfigure}[t]{0.32\linewidth}
    \includegraphics[width=\linewidth]{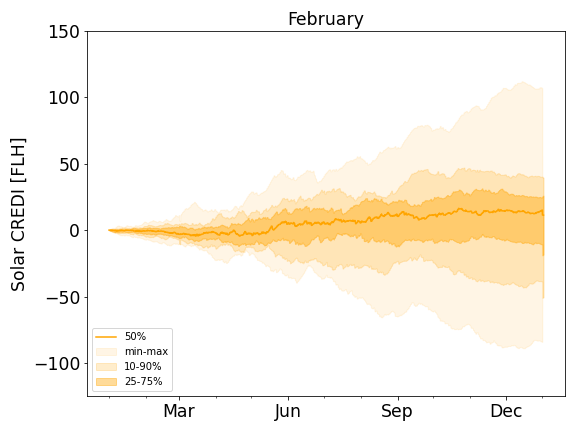}
    \caption{Febuary }
\end{subfigure}
\begin{subfigure}[t]{0.32\linewidth}
    \includegraphics[width=\linewidth]{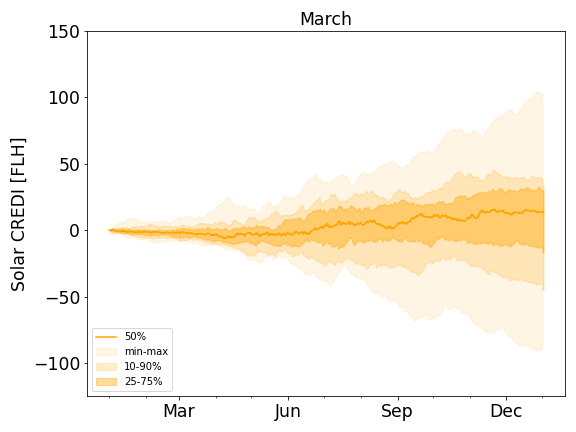}
    \caption{March }
\end{subfigure}
\begin{subfigure}[t]{0.32\linewidth}
    \includegraphics[width=\linewidth]{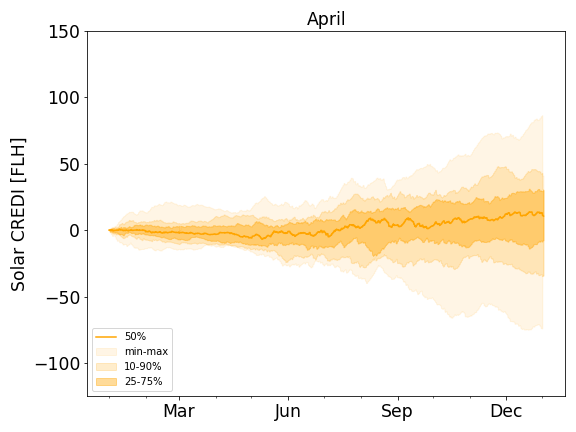}
    \caption{April }
\end{subfigure}
\begin{subfigure}[t]{0.32\linewidth}
    \includegraphics[width=\linewidth]{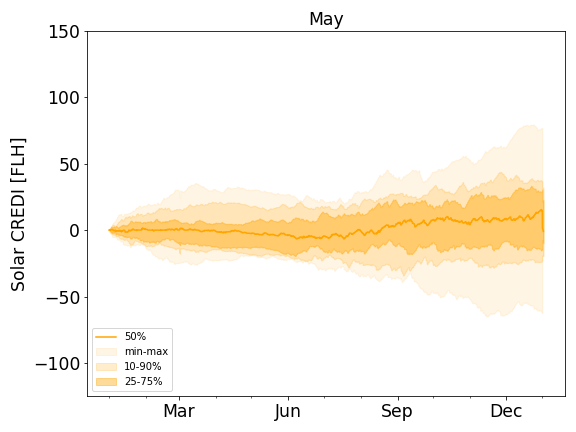}
    \caption{May }
\end{subfigure}
\begin{subfigure}[t]{0.32\linewidth}
    \includegraphics[width=\linewidth]{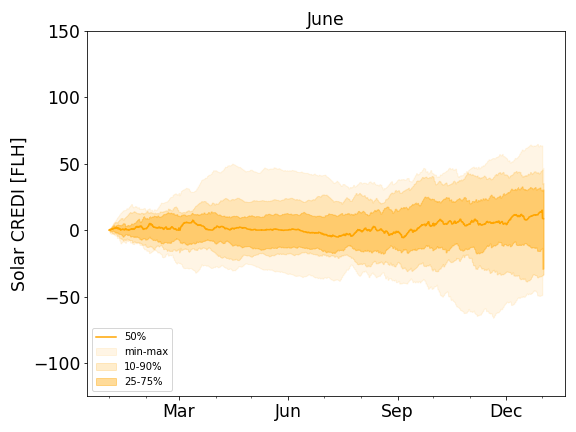}
    \caption{June }
\end{subfigure}
\begin{subfigure}[t]{0.32\linewidth}
    \includegraphics[width=\linewidth]{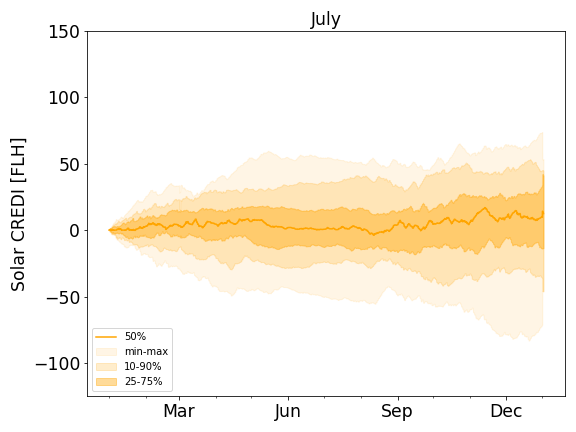}
    \caption{July }
\end{subfigure}
\begin{subfigure}[t]{0.32\linewidth}
    \includegraphics[width=\linewidth]{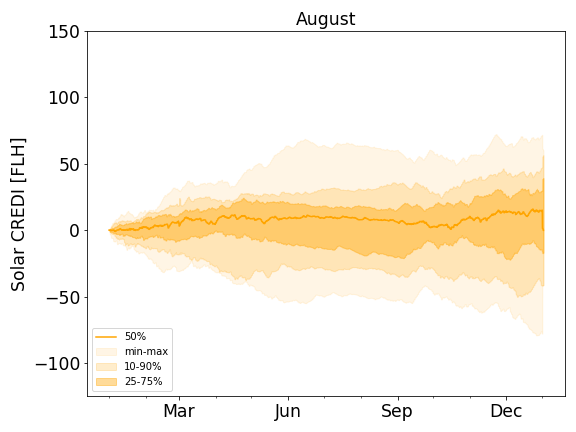}
    \caption{August }
\end{subfigure}
\begin{subfigure}[t]{0.32\linewidth}
    \includegraphics[width=\linewidth]{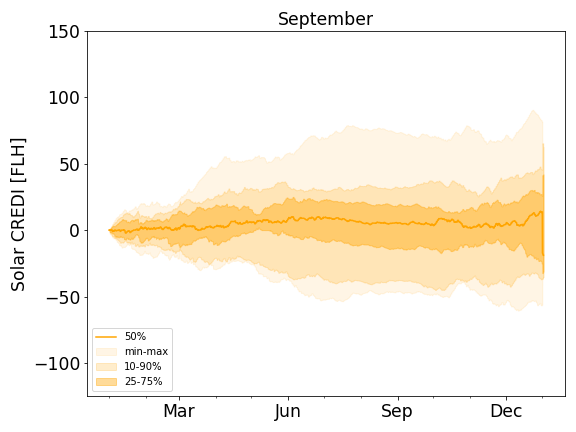}
    \caption{September }
\end{subfigure}
\begin{subfigure}[t]{0.32\linewidth}
    \includegraphics[width=\linewidth]{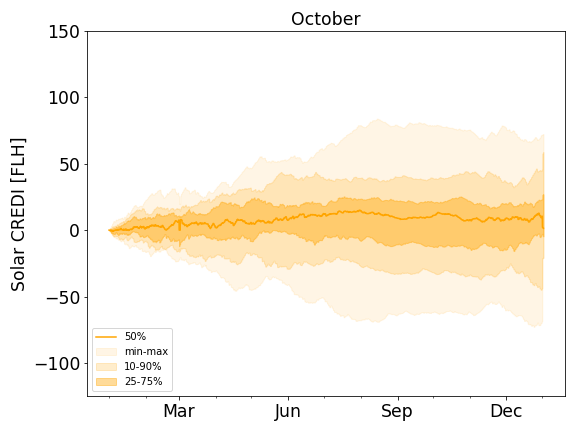}
    \caption{October }
\end{subfigure}
\begin{subfigure}[t]{0.32\linewidth}
    \includegraphics[width=\linewidth]{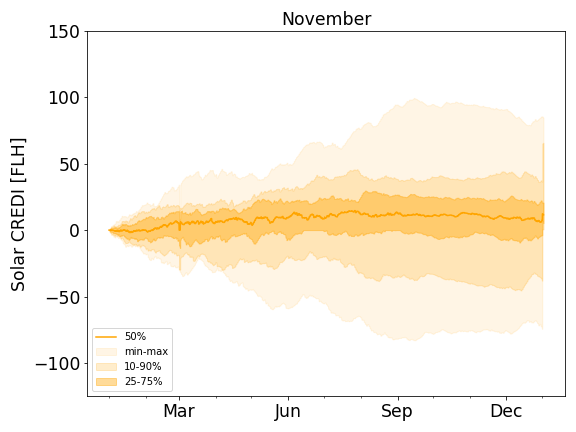}
    \caption{November}
\end{subfigure}
\begin{subfigure}[t]{0.32\linewidth}
    \includegraphics[width=\linewidth]{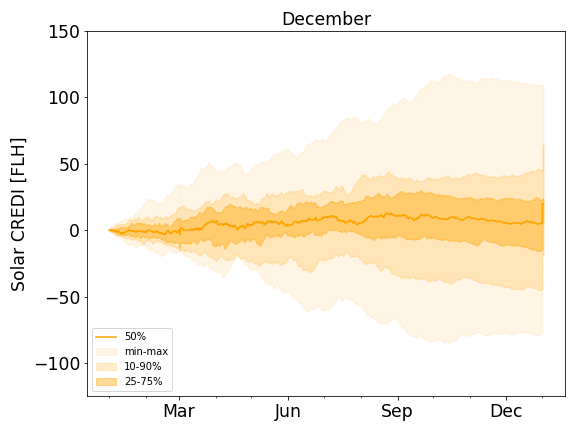}
    \caption{December }
\end{subfigure}
\caption{
    Comparison of the distribution of the \sdi{} with different the monthly starting points of the annual period. 
    The distribution is shown with the 50\ts{th} percentile (orange line), the 25-75, 10-90 percentile and min-max range (shaded orange, see legend) for each hour of the year for the years 1991-2020 in the `NL1' region. 
    Four exemplary storylines are shown, namely 1996 (red), 1998 (green), 2003 (purple) and 2016 (black), see main text for details and analysis.}
\label{SIfig:startdate_solar}
\end{figure}

\begin{figure}[b]
\centering
\begin{subfigure}[t]{0.32\linewidth}
    \includegraphics[width=\linewidth]{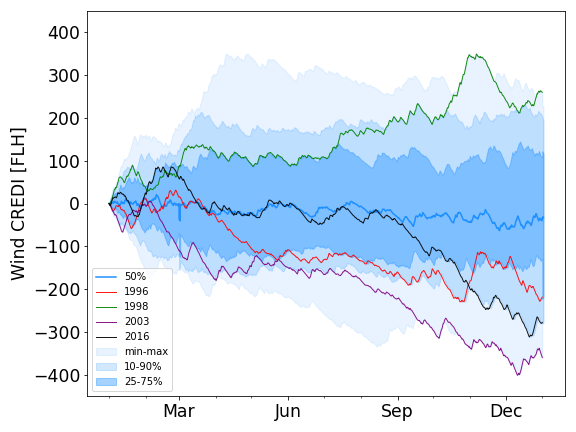}
    \caption{January }
\end{subfigure}
\begin{subfigure}[t]{0.32\linewidth}
    \includegraphics[width=\linewidth]{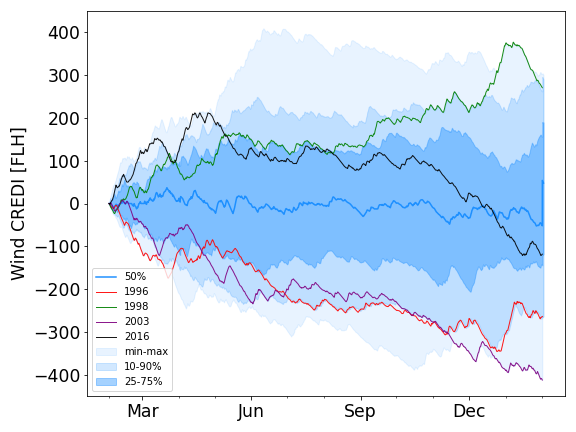}
    \caption{Febuary }
\end{subfigure}
\begin{subfigure}[t]{0.32\linewidth}
    \includegraphics[width=\linewidth]{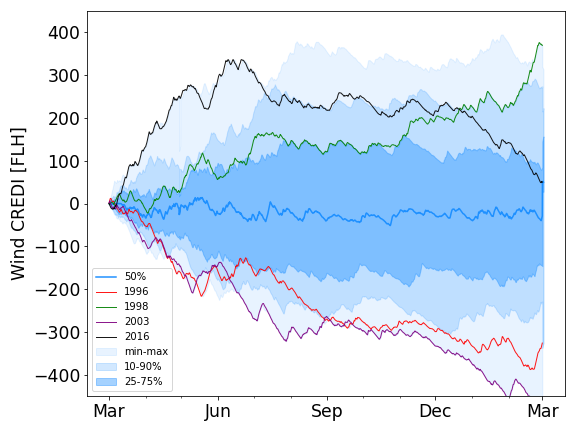}
    \caption{March }
\end{subfigure}
\begin{subfigure}[t]{0.32\linewidth}
    \includegraphics[width=\linewidth]{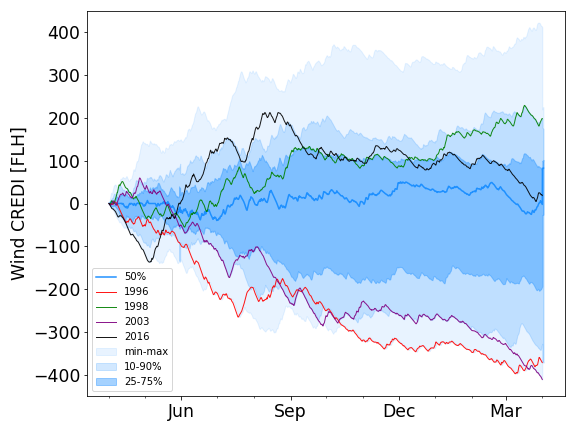}
    \caption{April }
\end{subfigure}
\begin{subfigure}[t]{0.32\linewidth}
    \includegraphics[width=\linewidth]{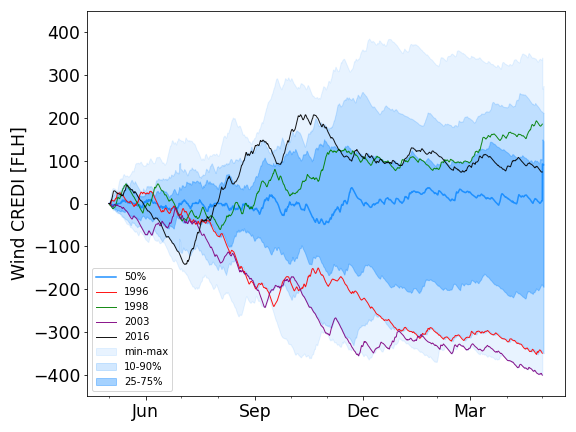}
    \caption{May }
\end{subfigure}
\begin{subfigure}[t]{0.32\linewidth}
    \includegraphics[width=\linewidth]{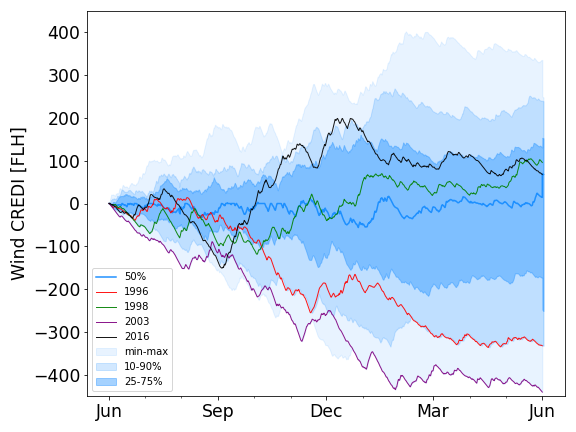}
    \caption{June }
\end{subfigure}
\begin{subfigure}[t]{0.32\linewidth}
    \includegraphics[width=\linewidth]{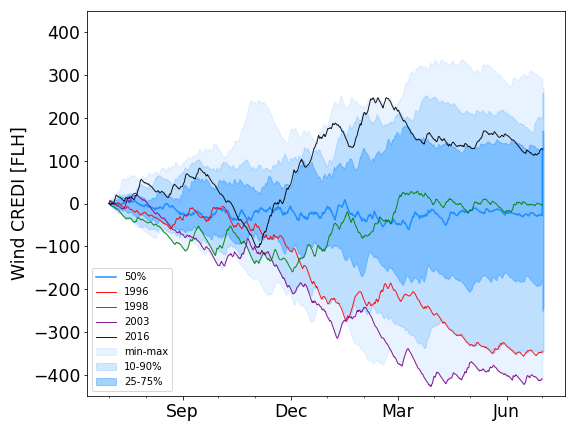}
    \caption{July }
\end{subfigure}
\begin{subfigure}[t]{0.32\linewidth}
    \includegraphics[width=\linewidth]{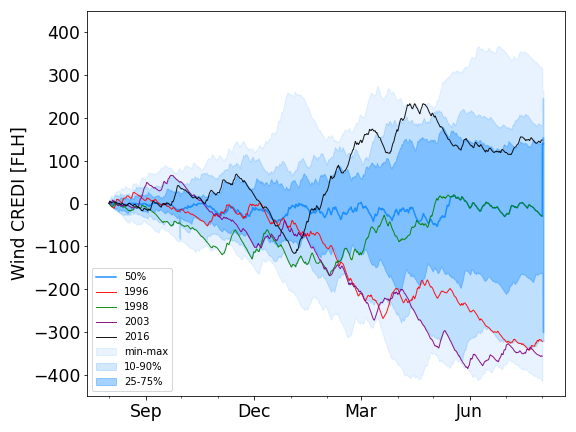}
    \caption{August }
\end{subfigure}
\begin{subfigure}[t]{0.32\linewidth}
    \includegraphics[width=\linewidth]{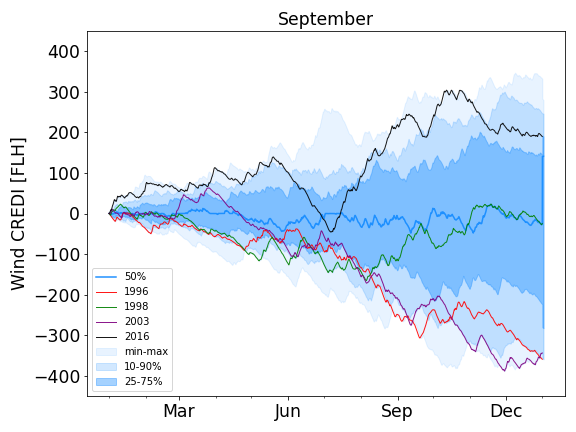}
    \caption{September }
\end{subfigure}
\begin{subfigure}[t]{0.32\linewidth}
    \includegraphics[width=\linewidth]{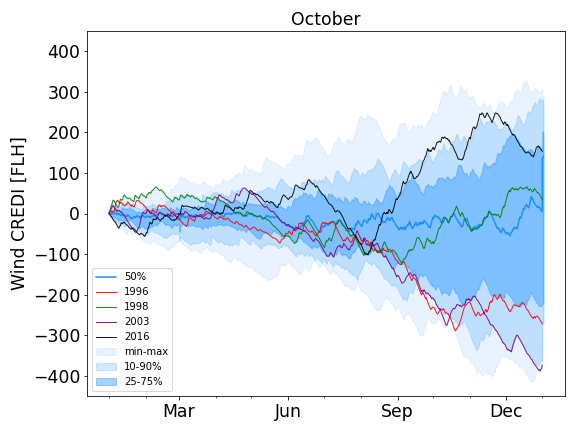}
    \caption{October }
\end{subfigure}
\begin{subfigure}[t]{0.32\linewidth}
    \includegraphics[width=\linewidth]{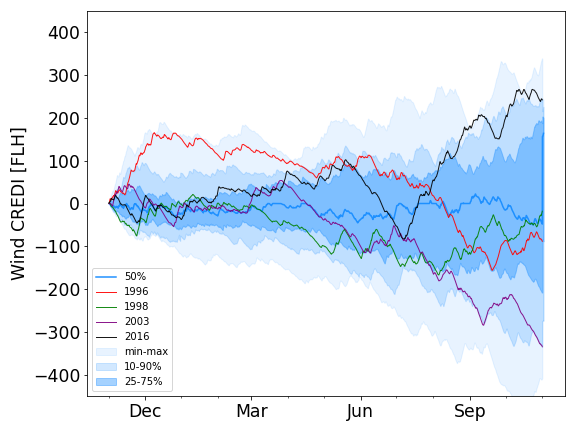}
    \caption{November}
\end{subfigure}
\begin{subfigure}[t]{0.32\linewidth}
    \includegraphics[width=\linewidth]{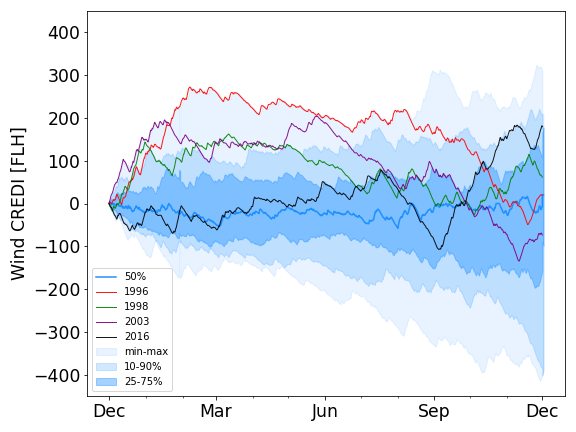}
    \caption{December }
\end{subfigure}
\caption{
    Comparison of the distribution of the \wdi{} with different the monthly starting points of the annual period. 
    The distribution is shown with the 50\ts{th} percentile (blue line), the 25-75, 10-90 percentile and min-max range (shaded blue, see legend) for each hour of the year for the years 1991-2020 in the `NL1' region. 
    Four exemplary storylines are shown, namely 1996 (red), 1998 (green), 2003 (purple) and 2016 (black), see main text for details and analysis.}
\label{SIfig:startdate_wind}
\end{figure}

Section 4 in the main text describes the application of the \credi{} at different timescales. 
Here we show how the hourly distribution of \credi{} changes over a year if a different starting point is used (Figures~\ref{SIfig:startdate_solar} \& \ref{SIfig:startdate_wind}). 
In line with the main text four exemplary storylines are shown, namely 1996 (red), 1998 (green), 2003 (purple) and 2016 (black).

From Figure~\ref{SIfig:startdate_wind}, the impact of choosing a different starting point becomes very clear.  
For the storylines shown you can see that they change from one of the highest, to on of the lowest depending on the start point. 
To a lesser degree, the same holds for the solar resource shown in  Figure~\ref{SIfig:startdate_solar}.

\clearpage
\section{Additional seasonal analysis figures of \credi}\label{SIC:seasonal}
\begin{figure}[h]
    \centering
    \includegraphics[width=\textwidth]{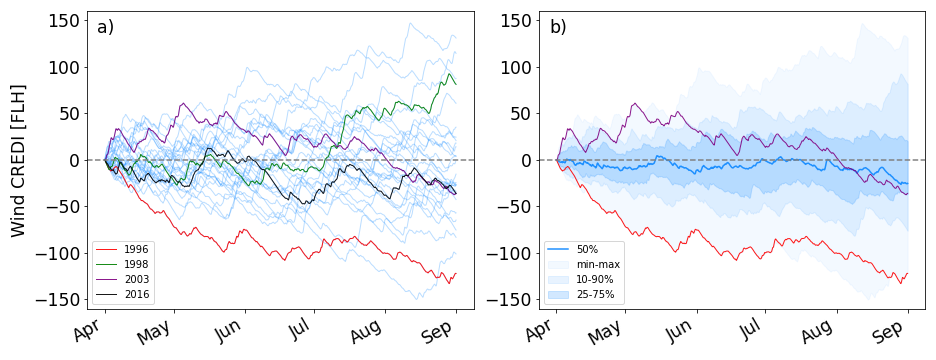}
    \caption{
    Hourly summer \wdi{} throughout the season over the period 1991-2020 for `NL1'.
    Figure a) shows the specific progression of \wdi{} for each summer season (blue lines).
    In addition, four example storylines are represented, namely 1996 (red), 1998(green), 2003(purple) and 2016 (black), see main text for details and analysis.
    Figure b) shows two storylines (1996, 2003) and the hourly distribution of the \wdi{}, namely the 50\ts{th} percentile (blue line), the 25-75, 10-90 percentile, and min-max range (shaded blue, see legend). 
    }
    \label{SIfig:analysis_season-summer_wind}
\end{figure}

\begin{figure}[b!]
    \centering
    \includegraphics[width=\textwidth]{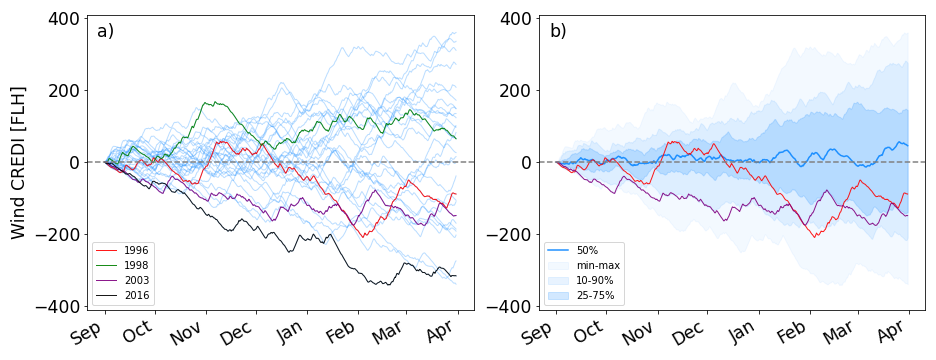}
    \caption{
    Hourly winter \wdi{} throughout the season over the period 1991-2020 for `NL1'.
    As shown in Figure~\ref{SIfig:analysis_season-summer_wind}.
    }
    \label{SIfig:analysis_season-winter_wind}
\end{figure}

\begin{figure}[ht!]
    \centering
    \includegraphics[width=\textwidth]{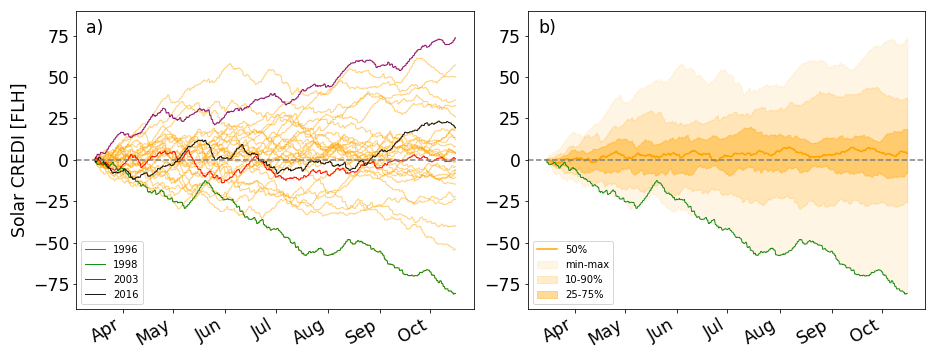}
    \caption{
    Hourly summer \sdi{} throughout the season over the period 1991-2020 for `NL1'.
    Figure a) shows the specific progression of \sdi{} for each summer season (orange lines).
    In addition, four example storylines are represented, namely 1996 (red), 1998(green), 2003(purple) and 2016 (black), see main text for details and analysis.
    Figure b) shows two storylines (1996, 2003) and the hourly distribution of the \sdi{}, namely the 50\ts{th} percentile (orange line), the 25-75, 10-90 percentile, and min-max range (shaded orange, see legend). 
    }
    \label{SIfig:analysis_season-summer_solar}
\end{figure}

\begin{figure}[h!]
    \centering
    \includegraphics[width=\textwidth]{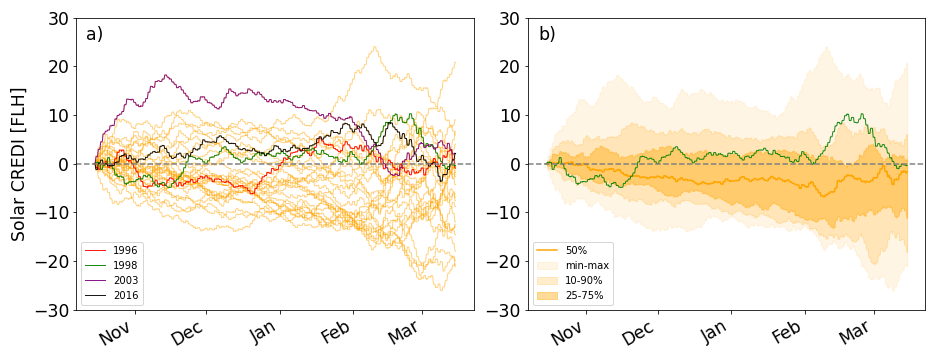}
    \caption{
    Hourly winter \sdi{} throughout the season over the period 1991-2020 for `NL1'.
    As shown in Figure~\ref{SIfig:analysis_season-summer_solar}.
    }
    \label{SIfig:analysis_season-winter_solar}
\end{figure}

Section 4.3 in the main text shows the seasonal variability in \credi.
Here we provide some additional figures representing a different season for either {\sc wind} or \sdi{} (Figures~\ref{SIfig:analysis_season-summer_wind}-\ref{SIfig:analysis_season-winter_solar}). 




\clearpage
\section{Additional short-term analysis figures of \credi}\label{SID:shortterm}
\begin{figure}[h]
        \centering
        \includegraphics[width=\textwidth]{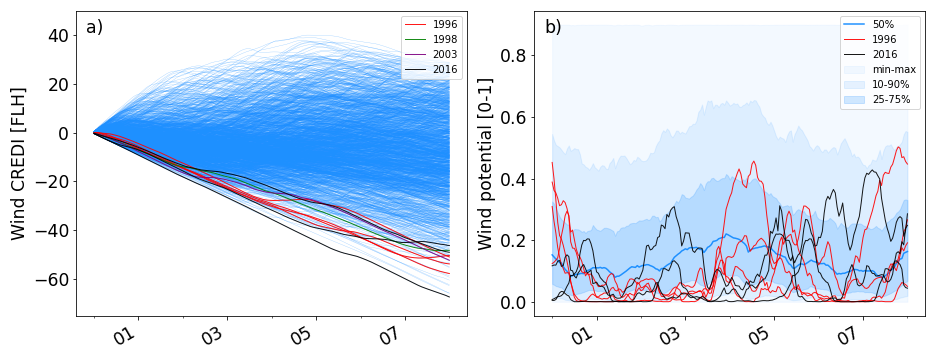}
        \caption{
                Hourly winter \wdi{} per 8-days for all events with less then 5 days overlapping in the period May 1991 to April 2021 for `NL1'. 
                The storylines show the analysis years 1996 (red, 4x), 1998 (green, 1x), 2003 (purple, 1x) and 2016 (black, 3x). 
                Figure a) shows the specific progression of \wdi{} for each summer season (blue lines). 
                To highlight the behaviour during an event, Figure b) shows the hourly distribution of the wind generation potential, namely the 50\ts{th} percentile (blue line), the 25-75, 10-90 percentile, and min-max range (shaded blue, see legend).
        }
        \label{SIfig:analysis_short-term-winter_wind_behaviour}
\end{figure}

Section 4.4 in the main text shows an example of the short-term \credi{} event selection.
Here we provide some additional figures related to the event selection and the observed behaviour.  
Figure~\ref{SIfig:analysis_short-term-winter_wind_behaviour} shows the wind distribution of the generation potential during the analysis period and for the selected events.

\begin{figure}[h]
        \centering
        \includegraphics[width=\textwidth]{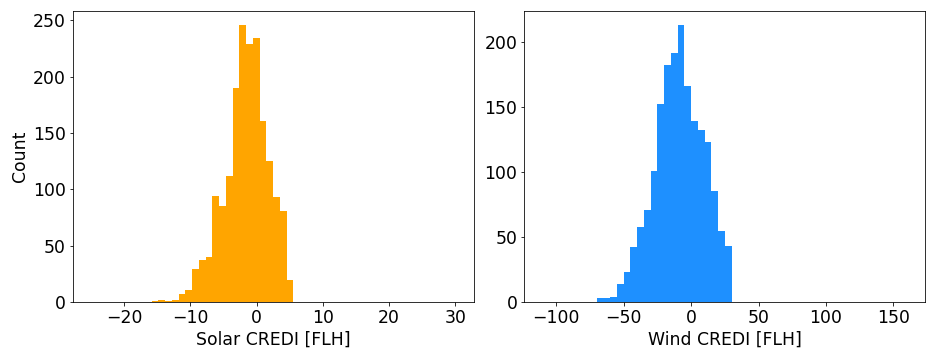}
        \caption{
                Histogram of the \sdi{} (Figure a) and \wdi{} (Figure b) at 8-days for all events with less then 5 days overlapping in the period May 1991 to April 2021 for `NL1'. 
        }
        \label{SIfig:analysis_short-term-winter_wind_histogram}
\end{figure}

\begin{figure}[h]
        \centering
        \includegraphics[width=0.6\textwidth]{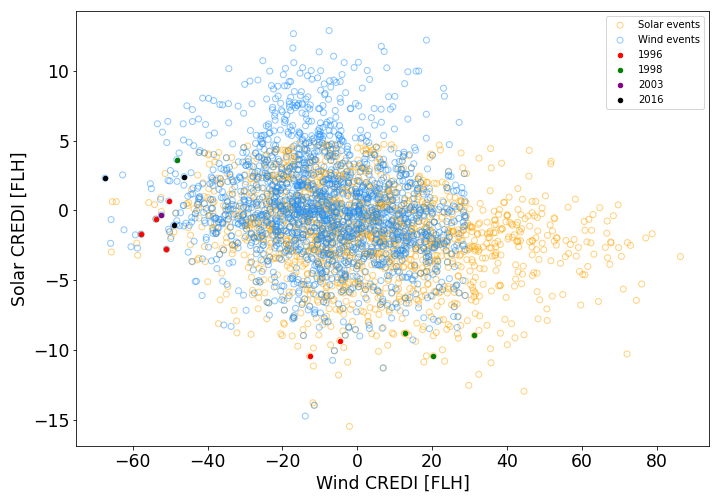}
        \caption{
                The 8-days \wdi{} and associated \sdi{} for all \wdi{} (blue) and \sdi{} (orange) events with less then 5 days overlapping in the period May 1991 to April 2021 for `NL1'. 
                The highlighted events are for those used in the analysis 1996 (red, 4x), 1998 (green, 1x), 2003 (purple, 1x) and 2016 (black, 3x). 
        }
        \label{SIfig:analysis_short-term-winter_wind_scatter}
\end{figure}

\begin{figure}[h]
        \centering
        \includegraphics[width=0.6\textwidth]{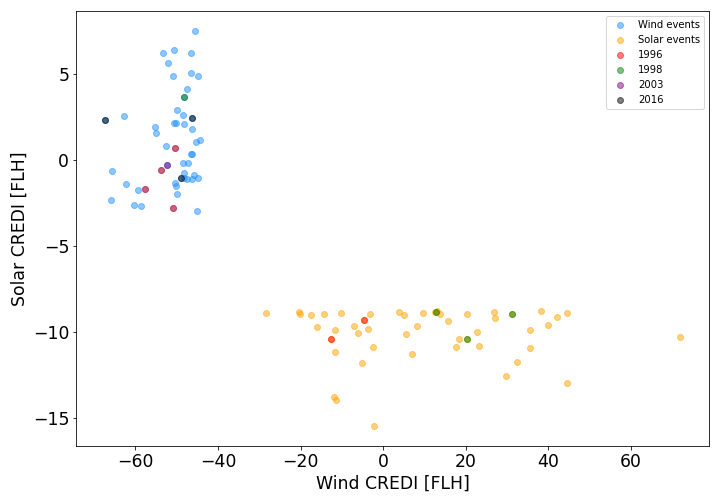}
        \caption{
                As Figure \ref{SIfig:analysis_short-term-winter_wind_scatter}, but then only the top 50 \wdi{} and \sdi{} events are shown. 
        }
        \label{SIfig:analysis_short-term-winter_wind_scatter_top50}
\end{figure}

The distribution of all non-overlapping events in the analysed period for both \sdi{} and \wdi is shown in Figure~\ref{SIfig:analysis_short-term-winter_wind_histogram}.
This is then further detailed by looking at the \wdi{} and \sdi{} values at the end of the selected events for both wind and solar in Figures~\ref{SIfig:analysis_short-term-winter_wind_scatter} and \ref{SIfig:analysis_short-term-winter_wind_scatter_top50}, where the latter only shows the top 50 events for both wind and solar.
A Table with all top 50 events for both {\sc wind} and \sdi{} 8-day events is provided in Table~\ref{tab:Top50longtable}, see the \emph{Open Research} section for the details of the code repository that contains the full list of all events.


\begin{longtable}{|r|rl|rl|}
\caption{
    Overview of the index value and event date for the top 50 8-day events selected both for {\sc wind} and \sdi{}. 
    Only those events are selected which have less then 5-days of overlap. 
    The full list of the all 8-day events can be found as listed in the \emph{Open Research} section.} \label{tab:Top50longtable} \\

\hline 
Event Rank & \sdi{} & Event date & \wdi{} & Event date  \\ \hline \hline
\endfirsthead

\multicolumn{5}{c}{{\bfseries \tablename\ \thetable{} -- continued from previous page}} \\ \hline
Event Rank & \sdi{} & Event date & \wdi{} & Event date  \\ \hline \hline
\endhead

\hline \multicolumn{5}{|c|}{{Continued on next page}} \\ \hline
\endfoot

\hline \hline
\endlastfoot

1         & -15,49          & 23/05/2013     & -67,36          & 24/01/2017     \\
2         & -13,98          & 18/05/1996     & -65,88          & 27/12/2006     \\
3         & -13,80          & 07/06/2012     & -65,75          & 30/12/1992     \\
4         & -12,97          & 06/05/2002     & -62,68          & 18/01/2013     \\
5         & -12,55          & 08/07/2002     & -62,30          & 30/01/1991     \\
6         & -11,82          & 26/05/2016     & -60,45          & 15/02/1993     \\
7         & -11,75          & 11/07/2020     & -59,27          & 01/02/1992     \\
8         & -11,30          & 18/06/1995     & -58,73          & 06/02/2006     \\
9         & -11,16          & 23/05/1994     & -57,74          & 13/12/1996     \\
10        & -10,93          & 28/05/2006     & -55,27          & 21/01/2001     \\
11        & -10,89          & 14/05/2010     & -54,99          & 15/12/2001     \\
12        & -10,88          & 28/07/2005     & -53,80          & 31/01/1997     \\
13        & -10,82          & 31/07/1993     & -53,39          & 18/02/2008     \\
14        & -10,44          & 21/03/1997     & -52,57          & 22/12/2007     \\
15        & -10,44          & 31/07/2011     & -52,30          & 28/01/2004     \\
16        & -10,43          & 13/06/1998     & -52,03          & 26/02/1994     \\
17        & -10,30          & 17/03/2019     & -50,96          & 26/01/1997     \\
18        & -10,12          & 12/05/2012     & -50,88          & 04/01/1993     \\
19        & -10,07          & 09/04/1993     & -50,61          & 02/03/2019     \\
20        & -10,00          & 12/05/2014     & -50,54          & 26/01/2019     \\
21        & -9,92           & 13/08/1993     & -50,41          & 03/02/2001     \\
22        & -9,88           & 14/06/2010     & -50,34          & 13/01/1997     \\
23        & -9,84           & 22/07/1993     & -50,13          & 13/01/2002     \\
24        & -9,74           & 26/03/2016     & -50,08          & 07/03/2021     \\
25        & -9,67           & 09/05/2010     & -49,91          & 17/02/1991     \\
26        & -9,65           & 05/04/2000     & -49,82          & 24/11/2011     \\
27        & -9,63           & 20/07/2011     & -49,03          & 21/12/2016     \\
28        & -9,39           & 13/07/2000     & -48,58          & 11/01/2003     \\
29        & -9,34           & 01/07/1996     & -48,47          & 14/12/2004     \\
30        & -9,17           & 03/07/1991     & -48,29          & 24/12/2021     \\
31        & -9,14           & 21/04/1992     & -48,21          & 22/11/1998     \\
32        & -9,01           & 06/05/2005     & -48,19          & 23/12/2017     \\
33        & -9,00           & 20/06/1993     & -48,14          & 16/10/1994     \\
34        & -8,99           & 10/09/1995     & -47,60          & 03/11/1997     \\
35        & -8,97           & 03/05/2019     & -47,44          & 05/12/2004     \\
36        & -8,97           & 31/05/1998     & -47,20          & 04/12/1991     \\
37        & -8,97           & 17/07/1998     & -46,51          & 07/02/2015     \\
38        & -8,95           & 31/03/2015     & -46,47          & 26/01/2015     \\
39        & -8,95           & 31/05/2014     & -46,44          & 04/02/1991     \\
40        & -8,93           & 02/06/2007     & -46,38          & 09/12/1991     \\
41        & -8,92           & 17/06/1991     & -46,31          & 09/01/2010     \\
42        & -8,89           & 08/03/2012     & -46,30          & 23/02/2013     \\
43        & -8,88           & 25/05/2003     & -46,24          & 29/01/2017     \\
44        & -8,86           & 14/08/2001     & -45,76          & 13/01/2013     \\
45        & -8,85           & 20/04/2005     & -45,51          & 18/02/2003     \\
46        & -8,84           & 01/07/2017     & -45,25          & 16/01/2001     \\
47        & -8,82           & 30/09/1991     & -45,16          & 17/03/1991     \\
48        & -8,82           & 09/10/1998     & -44,90          & 24/02/2018     \\
49        & -8,81           & 26/07/2011     & -44,87          & 27/01/2010     \\
50        & -8,80           & 08/05/1991     & -44,46          & 17/10/1995     
\end{longtable}  


\clearpage
\section{Application of \credi{} to other regions}\label{SIE:regions}
Section 5 in the main text discusses the use of the \credi{} for other regions. 
Here selected additional figures on the application of the index and very limited analysis for some other regions is provided.
Due to the preliminary version of the PECDv4.0 used, caution is advised on the exact interpretation of the results and no data is  provided for these regions. 
In addition, for the analysis only the seasonal and annual to decadal variability is discussed as the analysis of the short-term and sub-seasonal variability depends on the choice of the storylines which depend on the region considered and are kept consistent with the main text for reference.

The additional regions used are Slovakia (`SK00'), the southern tip of Sweden (`SE02') and one of the south-east regions of France (`FR10').
The choice for these regions was made to reflect some of the diversity within Europe.
Not all regions are shown for all figures provided in the main text, the figures not shown can be found as listed in the \emph{Open Research} section on github.

\subsection{Observed variability of wind and solar energy potential --- Other regions}
\begin{figure}[hb]
\centering
    \begin{subfigure}[t]{\linewidth}
        \centering
        \includegraphics[width=\textwidth]{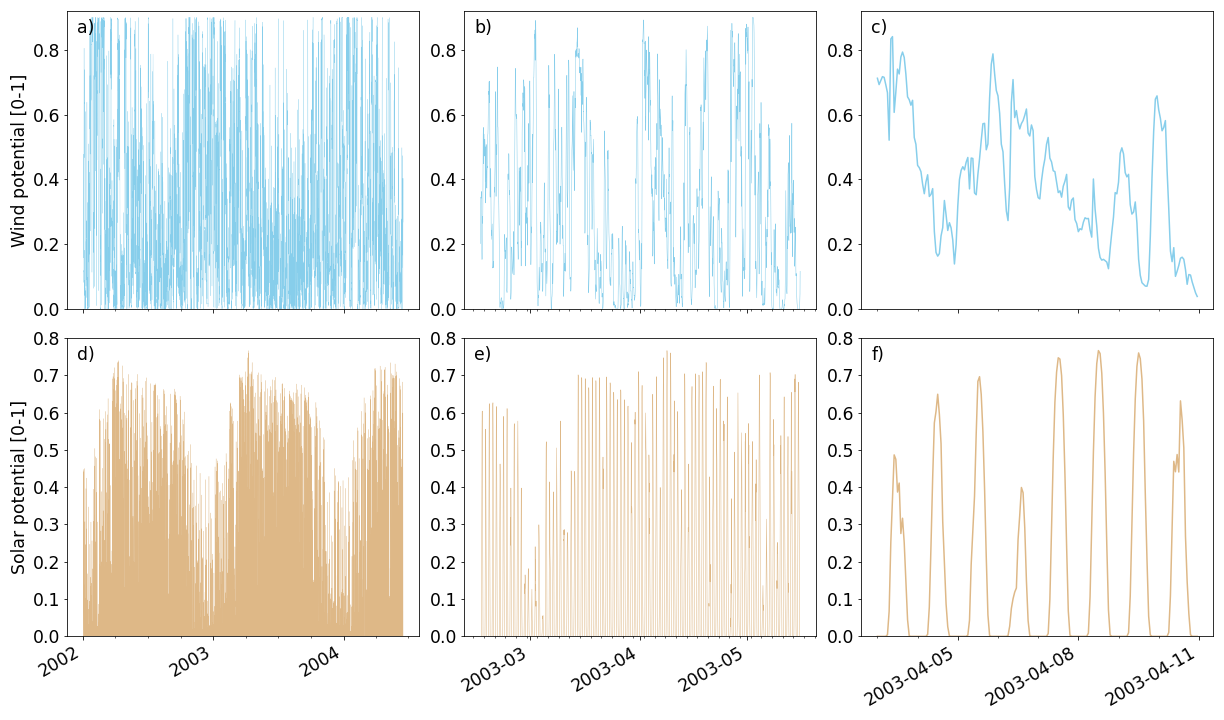}
        \caption{South-East of France (`FR10')}
    \end{subfigure}
    \begin{subfigure}[t]{\linewidth}
        \centering
        \includegraphics[width=\textwidth]{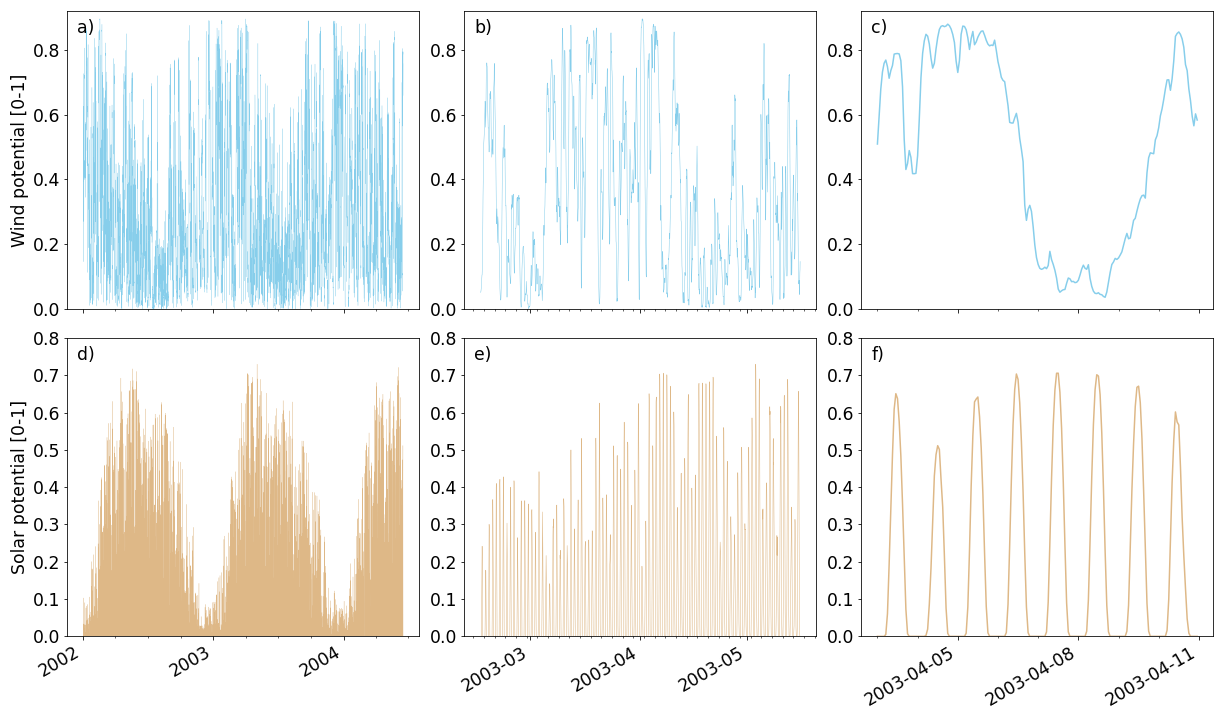}
        \caption{Southern tip of Sweden (`SE02')}
    \end{subfigure}
    \caption{
        As Figure~2 in the main text, but then for the regions as listed for 2002-2004.
        Timeseries of hourly generation potential of wind (top) and solar (bottom). 
        Showing variability on yearly (a,d), sub-seasonal (b,e) and daily (c,f) timescales. 
    }
    \label{SIfig:climatological_behaviour_other-regions}
\end{figure}

Similar observations can be made on the timescales of variability for the other regions then the `NL1' region discussed in the main text (Figure~\ref{SIfig:climatological_behaviour_other-regions}). 
While the distribution of the values differs between regions, similar characteristics are observed. 
For wind at seasonal timescales a lower mean generation potential is observed in all three regions (`SK00' not shown). 
Some shifts in the characteristic behaviour can be observed. 
For instance, there is lower solar generation in winter for more northern regions, and a more strongly pronounced skewness of the solar generation potential throughout the year is seen for `SE02'. 

\subsection{A hourly rolling windows climate --- Other regions}
\begin{figure}[hb]
    \centering
    \begin{subfigure}[t]{0.9\linewidth}
        \centering
        \includegraphics[width=\textwidth]{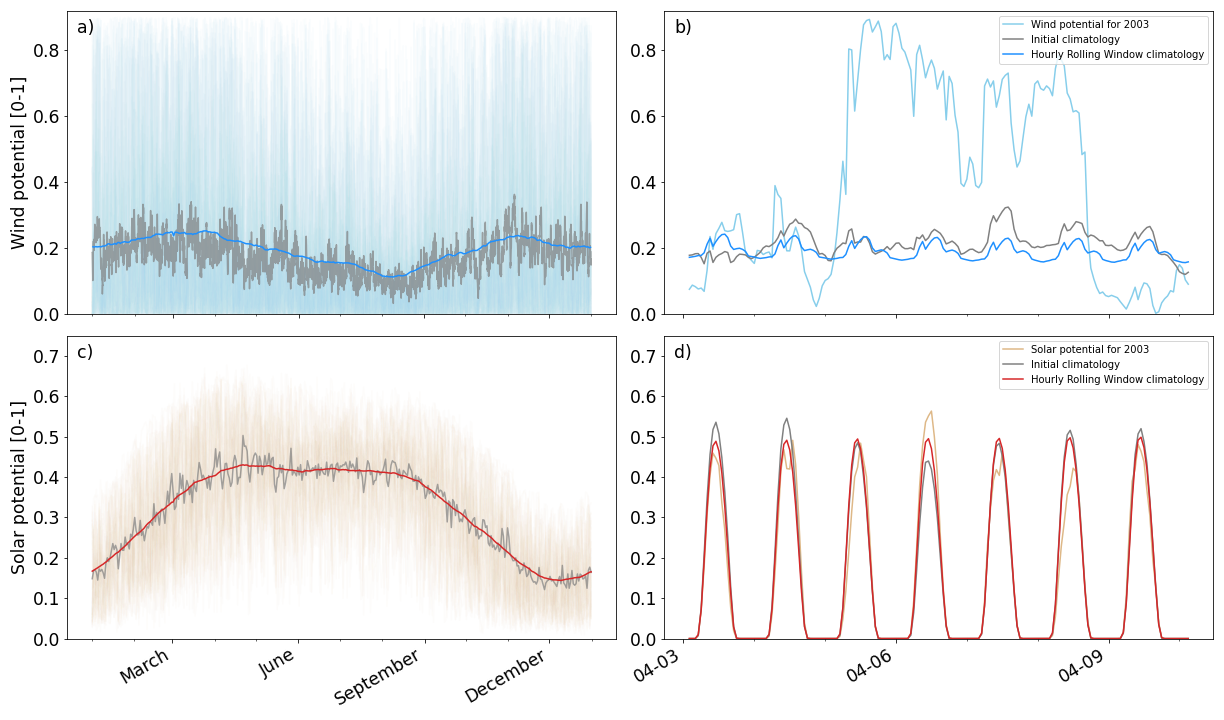}
        \caption{Slovakia (`SK00')}
    \end{subfigure}
    \begin{subfigure}[t]{0.9\linewidth}
        \centering
        \includegraphics[width=\textwidth]{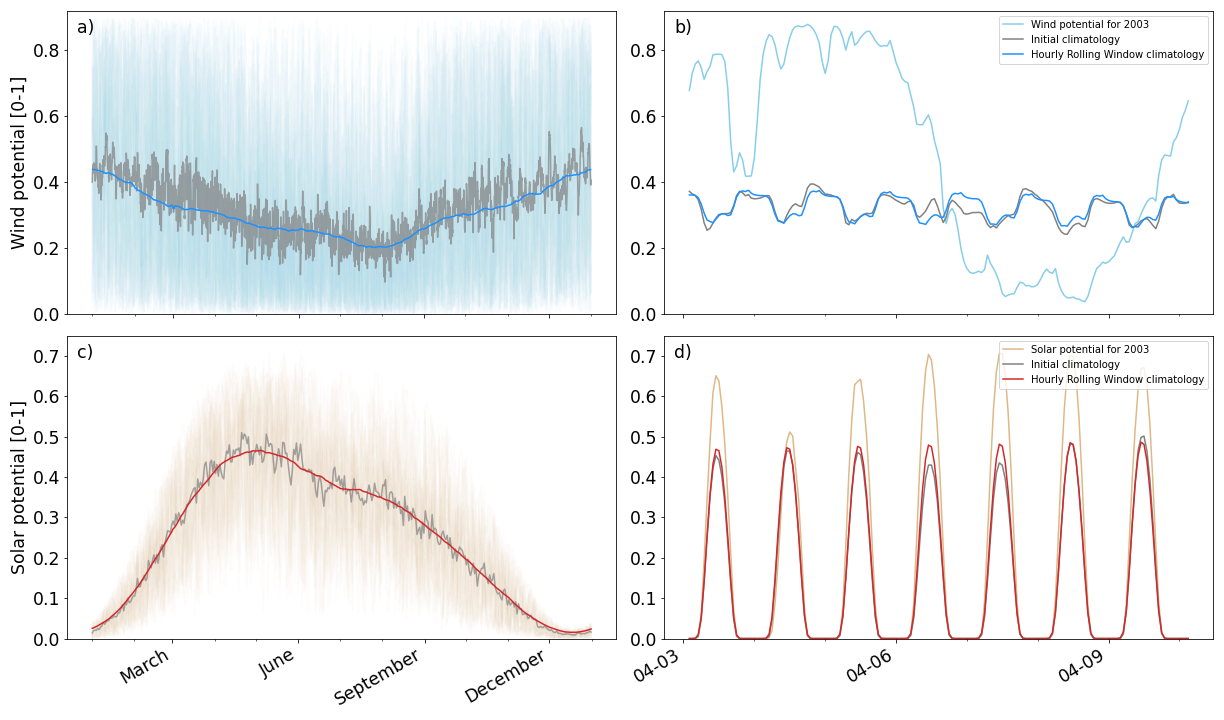}
        \caption{Southern tip of Sweden (`SE02')}
    \end{subfigure}
    \caption{
        As Figure~3 in the main text, but then for the regions as listed.
        Comparison of different methods for computing the climate of the potential generation for wind (top), and solar (bottom), for the period 1991-2020. 
        Figures (a,c) show the hourly generation potentials for each year in this period (light blue for wind and orange for solar), the initial, simple average-based climate (grey, see main text for details) and the hourly rolling window climate (blue and red, for wind, solar, respectively). 
        Figures (b,d) show the same, but specifically for the period 3-10 April 2003. 
        For clarity only 13:00 for each day of the year is shown in Figure (c).}
    \label{SIfig:climate_other-regions}
\end{figure}

The \emph{hourly rolling window} climate defined in Section~2.2 of the main text was applied without any changes to the other regions considered.
As can be seen in Figure~\ref{SIfig:climate_other-regions}, this climate provides a smoother description of the expected behaviour on annual timescales and reduces the random fluctuations. 

In line with the observations for the northern region of the Netherlands, the initial, simple average-based climate does capture the annual timescales, but shows random fluctuations from day-to-day and hour-to-hour. 
For both Slovakia and the part of Sweden shown some consistent daily variation is observed for their wind generation potential, whether this is from a physical driver is unknown and should be further studied before using the climatic description for these regions.

\clearpage
\subsection{Annual to decadal variability --- Other regions}
Section 4.1 in the main text discusses annual to decadal variability observed in the \credi{}, here we shortly discuss the same for other regions. 

\begin{figure}[hb]
    \centering
    \includegraphics[width=\textwidth]{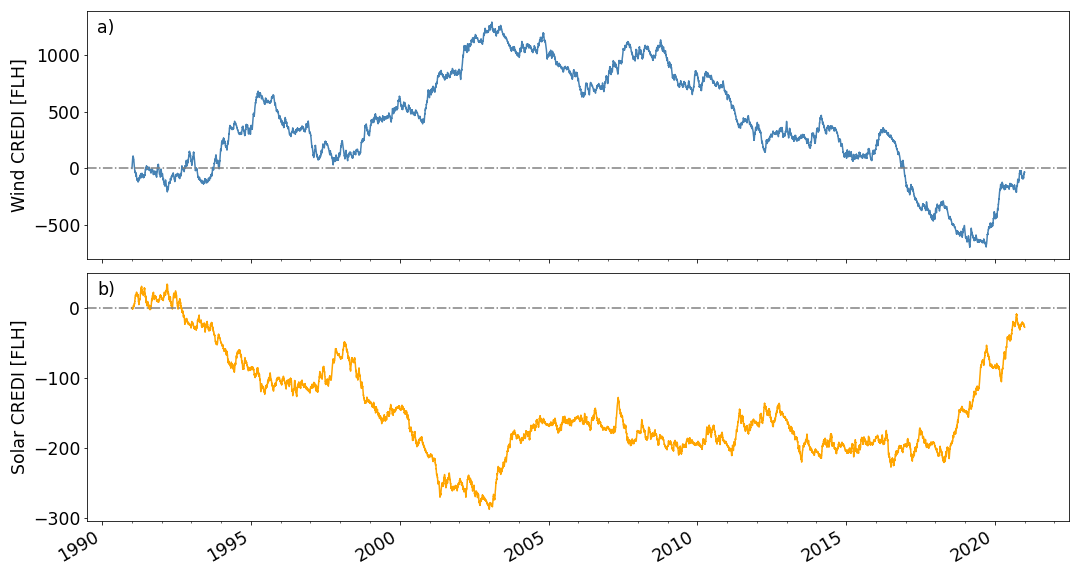}
    \caption{
        As Figure~4 in the main text, but then for the South-East of France (`FR10').
        Hourly Wind (a) and Solar (b) \credi{} over the period 1991-2020 for `NL1'. 
        As the climate was calculated over the same period, by definition the \credi{} sums to zero over the full period.}
    \label{SIfig:analysis_decadal_other-regions_A}
\end{figure}

\begin{figure}[ht]
    \centering
    \includegraphics[width=\textwidth]{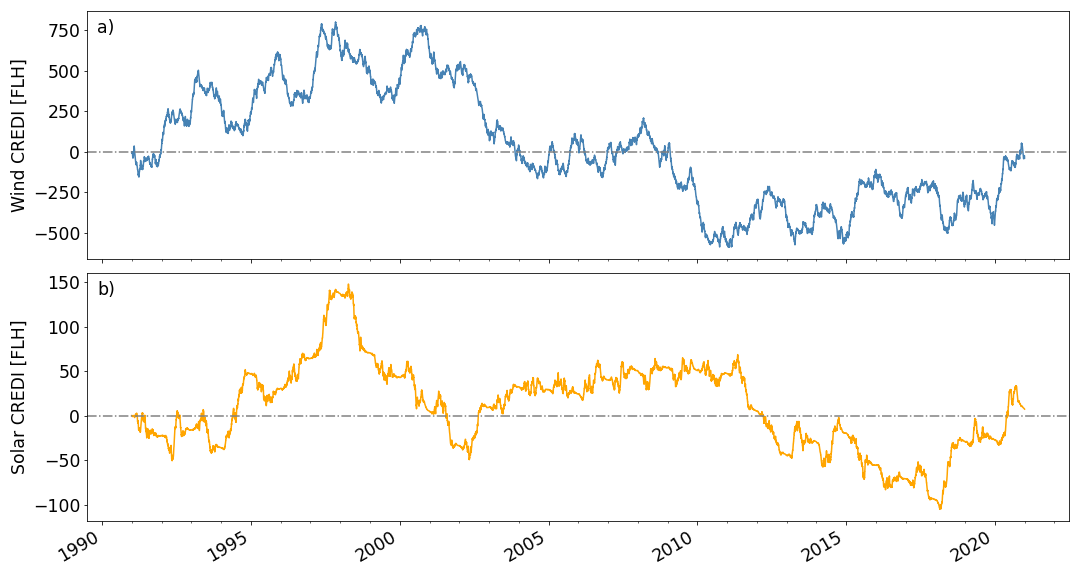}
    \caption{
        As Figure~4 in the main text, but then for the Southern tip of Sweden (`SE02') region. 
        Hourly Wind (a) and Solar (b) \credi{} over the period 1991-2020 for `NL1'. 
        As the climate was calculated over the same period, by definition the \credi{} sums to zero over the full period.}
    \label{SIfig:analysis_decadal_other-regions_B}
\end{figure}

Over the past 30~years, large and consistent inter-annual variation is observed in the \wdi{} for the `FR10' region (Figure~\ref{SIfig:analysis_decadal_other-regions_A}), while the `SE02' region shows more variable behaviour on annual and seasonal timescales (Figure~\ref{SIfig:analysis_decadal_other-regions_B}).
For the French region, some cumulative effect over the whole period can be observed, while the Swedish region shows a more oscillating pattern. 

Similar to the `NL1' region, more inter-annual periods with a flat \sdi{} can be observed then for wind. 
For the French region a general decrease of the \sdi, thus anomalous low generation potential, is observed in the period 1992-2004 and a very consistent increase from 2018 to 2021.
For the Swedish region a yearly flat \sdi{} is observed, likely related to the very limited solar generation potential in the winter. 

\newpage
\subsection{Seasonal variability --- Other regions}
\begin{figure}[hb]
    \centering
    \begin{subfigure}[t]{\linewidth}
        \centering
        \includegraphics[width=\textwidth]{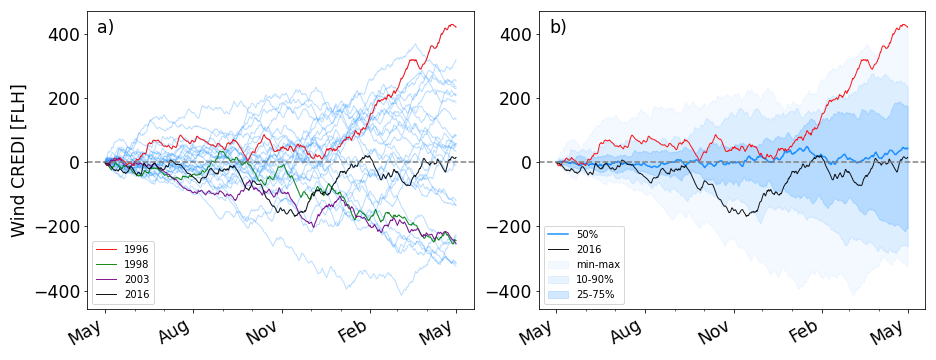}
        \caption{\wdi{}}\vspace*{0.5cm}
    \end{subfigure}
    
    \begin{subfigure}[t]{\linewidth}
        \centering
        \includegraphics[width=\textwidth]{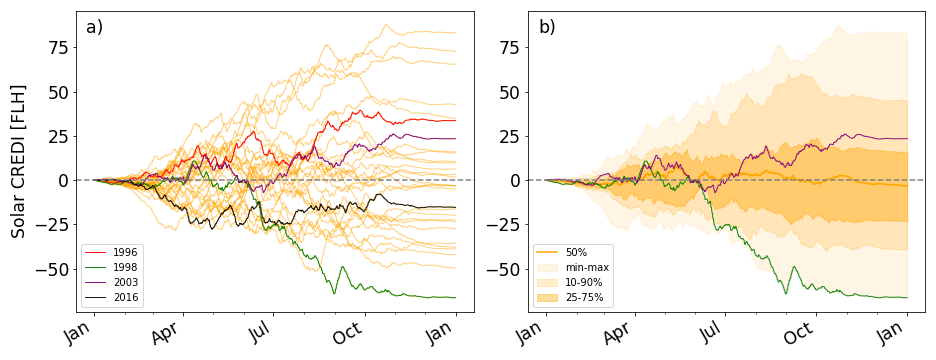}
        \caption{\sdi{}}\vspace*{0.5cm}
    \end{subfigure}
    \caption{
        As Figure~5 (here Figure~(a), blue shades) and 6 (here Figure~(b), orange shades) in the main text, but then for the Southern tip of Sweden (`SE02').
        Hourly \wdi{} per analysis year over the period May 1991 to April 2021 for `NL1'. 
        Figure a) shows the specific progression of \wdi{} for each year. 
        Figure b) shows the distribution of the \wdi{} for each hour of the year, namely the 50\ts{th} percentile, the 25-75, 10-90 percentile and min-max range (see legend). 
        Four exemplary storylines are shown, namely 1996 (red), 1998 (green), 2003 (purple) and 2016 (black).
    }
    \label{SIfig:analysis_seasonal_wind_other-regions}
\end{figure}

Section 4.2 in the main text discusses seasonal variability in the \credi{}, here we shortly discuss the same for the `SE02' region as it shows the most interesting properties (see Figure~\ref{SIfig:analysis_seasonal_wind_other-regions}). 

The \wdi{} in this Swedish region shows similar behaviour as the Dutch region discussed in the main text, but while the 2016 storyline is considered to be the most extreme for the northern region of the Netherlands, this is not the case for the `SE02' region. 
In addition, the shape of the distribution of the \wdi{} is different throughout the year and the 1996 storyline shows the highest \wdi{} value. 
This stark opposition to the behaviour observed in that storyline for the Netherlands indicates some possible balancing for this specific storyline.

The \sdi{} in the southern Swedish region `SE02' shows a very flat value in the period from October to March. 
This is likely due to the very clearly limited solar generation potential in this region during the wintertime period and the reasons for the limited annual to decadal variability observed for this region (see Figure~\ref{SIfig:analysis_decadal_other-regions_B}). 
At the same time large seasonal differences between the different March to September periods are observed. 
As for the `NL1' region, the year 1998 is the most extreme storyline for \sdi.

\clearpage
\section{Extended Data section}\label{SIF:dataext}
We used the preliminary 4th version of the Pan-European Climate Database to demonstrate the \credi{} in this paper~\parencite[PECDv4.0;][]{Dubus2022PECD}. 
This database, developed by Copernicus Climate Change Services (C3S) in cooperation with the European Network of Transmission System Operators for Electricity (ENTSO-E) will be the new standard database used for all common Transmission System Operator (TSO) studies. 
The full database will be openly available as part of the new C3S-Energy dataset, expected in late 2023 (\url{https://climate.copernicus.eu/energy/}). 
To showcase the developed index all figures show data from the preliminary PECDv4.0 of the northern region of the Netherlands (`NL1').

Within PECDv4.0 a range of technological properties have been modelled for both wind turbines and photovoltaic solar panels~\parencite{Dubus2022PECD}. 
Only the historic hourly generation potential (or capacity factor) timeseries are used for solar and wind with the properties of `existing technologies'.  
Our subset uses the ERA5 reanalysis for its meteorological forcing~\parencite{Hersbach2020}. 
The wind power plant conversion model is the generic power curve model presented in \textcite{Murcia2022} that is implemented in PyWake~\parencite{pywake}. 
For the property parameterisation it uses the 2020 data from the WindPowerNet (\url{https://www.thewindpower.net/}). 
Storm shut down behaviour is modelled after \textcite{MurciaLeon2021}, while wakes are modelled as part of the generic power curve and for other losses a 10\% reduction factor is applied~\parencite{Luzia2023}. 
The regional solar photo-voltaic (PV) potential is derived following \textcite{SaintDrenan2018}. 
A distribution of near optimal tilt and azimuth angles was used that reflects current installed capacities. 
For aggregation to the modelled zones in the PECDv4.0 database, the gridded ERA5 data was weighted by the cover of protected areas, regions with high slopes and/or high elevation.


\end{document}